\documentclass[preprint,showpacs,preprintnumbers,amsmath,amssymb,nofootinbib]{revtex4}
\usepackage{subfigure}
\usepackage{graphicx,textcomp}
\usepackage{dcolumn}
\usepackage{bm}
\usepackage{mathrsfs,slashed,amsmath}
\usepackage{natbib}
\usepackage{dsfont}
\allowdisplaybreaks
\newcommand{\pt}{$\bullet$}
\begin{document}
\title{Can Baryon Chiral Perturbation Theory be used to extrapolate lattice data for the moment $\langle x\rangle_{u-d}$ of the nucleon?}
\author{Peter C.~Bruns}
\affiliation{Institut f\"ur Theoretische Physik, Universit\"at Regensburg, D-93040 Regensburg, Germany}
\author{Ludwig Greil}
\affiliation{Institut f\"ur Theoretische Physik, Universit\"at Regensburg, D-93040 Regensburg, Germany}
\author{Rudolf R\"odl}
\affiliation{Institut f\"ur Theoretische Physik, Universit\"at Regensburg, D-93040 Regensburg, Germany}
\author{Andreas Sch\"afer}
\affiliation{Institut f\"ur Theoretische Physik, Universit\"at Regensburg, D-93040 Regensburg, Germany}
\date{\today}
\begin{abstract}
We discuss the question in the title employing manifestly covariant Baryon Chiral Perturbation Theory and recent high-statistics lattice results published in \cite{Bali:2014gha}.
\end{abstract}
\maketitle
\section{Introduction\label{sec:intro}}
Lattice Quantum Chromodynamics (LQCD) has now reached the point where fully dynamical results for nucleon structure observables are available for quark masses almost corresponding to the physical pion mass $M_{\pi}\approx 135\,\mathrm{MeV}$ (for recent reviews, see \cite{Renner:2010ks,Meyer:2011zh,Syritsyn:2014saa,Alexandrou:2014yha}). For some, but not all of these quantities, also results extracted from phenomenology exist, and it is crucial to compare the lattice results with the experimental results, in order to judge the reliablility of LQCD predictions for such quantities which are not easily accessible in experiments. For some observables, for example the axial coupling constant $g_{A}$ of the nucleon, and the moment $\langle x\rangle_{u-d}$ (isovector quark momentum fraction) of the nucleon, there is substantial tension between the experimental and the lattice outcome. For the latter quantity, which is the topic of the present article, we refer in particular to the very recent results and discussion in \cite{Bali:2014gha}.\\
The quark mass dependence of hadron observables is given by Chiral Perturbation Theory (ChPT) \cite{Weinberg:1978kz,Gasser:1983yg,Gasser:1984gg,Gasser:1987rb}, the low-energy effective field theory of QCD. It has often been used to extrapolate lattice data employing quark masses larger than in the real world down to the experimental point. Moreover, the dependence on the spatial volume $L^3$ of the lattice can also be calculated within the same theoretical framework. This was first demonstrated in the purely Goldstone-bosonic sector \cite{Gasser:1987zq,Hasenfratz:1989pk,Luscher:1983rk}, and later extended to the one-baryon sector \cite{AliKhan:2003cu,Khan:2006de,Beane:2004tw,Beane:2004rf,Detmold:2004ap,Detmold:2005pt}. We work in the so-called $p$-regime where the counting scheme is set up such that $M_{\pi}\sim L^{-1}\sim\mathcal{O}(p)$, where $p$ denotes a small quantity like a (pseudo-)Goldstone boson mass or momentum. It has become a common rule of thumb that one 
should have $M_{\pi}L\gtrsim 4$ in order to have the finite volume corrections under theoretical control in this regime \cite{Beane:2011pc,Geng:2011wq,Alvarez-Ruso:2013fza,Walker-Loud:2013yua} (for the case of the pion mass at two-loop accuracy, \cite{Colangelo:2006mp} gives a lower bound of $M_{\pi}L\gtrsim 2$ for the applicability of the $p$-regime). On the other hand, there is a long-standing debate in the literature about the range of applicability of the chiral extrapolation formulae given by (two-flavor) Baryon Chiral Perturbation Theory (BChPT) in terms of the quark masses or $M_{\pi}$, see \cite{Walker-Loud:2013yua,McGovern:1998tm,Young:2002ib,Bernard:2003rp,Beane:2004ks,Djukanovic:2006xc,McGovern:2006fm,Schindler:2006ha,Colangelo:2006mp,Bernard:2007zu,Hall:2010ai,Bali:2012qs,Bruns:2012eh,Beane:2014oea} for relevant references. The opinion most generally shared today seems to be that earlier applications of BChPT formulae to lattice results for pion masses $M_{\pi}\gtrsim 500\,\mathrm{MeV}$ were not 
under sufficient control for a reliable extrapolation to the physical 
point, while (citing the review \cite{Bernard:2007zu}) ``it is fair to say that chiral extrapolations of nucleon properties can be trusted for pion masses below $\sim 350\,\mathrm{MeV}$``. Of course, such statements will also depend on the particular channel or observable one is considering, on the chiral order (accuracy) of the calculation, and on the available information on relevant low-energy constants (LECs) from other sources. Moreover, in a finite spatial volume $L^3$, the pion masses should not be smaller than $\sim L^{-1}$ in order to stay in the above-mentioned $p$-regime.\\In this work, we will study the chiral extrapolation and the finite volume corrections of lattice results for the moment $\langle x\rangle_{u-d}$ of the nucleon (for the pertinent definitions and conventions, see e.~g.~\cite{Ji:1998pc,Diehl:2003ny,Renner:2010ks,Bali:2014gha,Dorati:2007bk,Wein:2014wma}). First, in Sec.~\ref{sec:stab}, we neglect the finite volume corrections, and use a truncated version of the full one-loop 
expression for $\langle x\rangle_{u-d}(M_{\pi})$ to test the stability of the chiral fits with respect to a variation of supplemented terms of leading two-loop order ($\mathcal{O}(p^4)$). Encouraged by this first test, we apply the formulae derived in \cite{Dorati:2007bk,Wein:2014wma,Greil:2014awa} (in the general setting of manifestly covariant BChPT \cite{Becher:1999he}) to a more complete data set, and examine the finite volume corrections together with the chiral extrapolation in the pion mass in Sec.~\ref{sec:finite}, in three different fit scenarios. The plots in Fig.~\ref{fig:fvfits} illustrate our main results, corresponding to the fit parameters tabulated in Tab.~\ref{tab:fvfits}. We discuss our findings in Sec.~\ref{sec:disc}, from two perspectives:
\begin{itemize}
 \item[\pt] The standard perspective, most commonly found in the literature, where lattice data at unphysically large quark masses is connected to the region of low quark masses by means of a chiral extrapolation, and compared with some known experimental value, and
 \item[\pt] from a perspective where it is assumed that the experimental value is unknown, and where ChPT is applied to test the internal consistency of the lattice data, in particular for quark masses close to the physical values.
\end{itemize}
With the example of $\langle x\rangle_{u-d}$ at hand, we shall demonstrate below that, from the second perspective, BChPT can play an important role for future lattice simulations, as an indicator of the possible presence of uncontrolled systematic errors inherent in the data, even if there are data points at (nearly) physical quark masses and large volumes.  
Finally, we try to give a well-founded answer to the question posed in the title in our concluding Sec.~\ref{sec:conc}.
\section{Stability test of chiral fits for $\langle x\rangle_{u-d}$\label{sec:stab}}
The chiral expansion of $\langle x\rangle_{u-d}$ to $\mathcal{O}(p^4)$ in BChPT reads \cite{Chen:2001pva,Belitsky:2002jp,Arndt:2001ye,Ando:2006sk,Diehl:2006js,Dorati:2007bk,Moiseeva:2012zi,Wein:2014wma}
\begin{align}
\begin{split}
\langle x\rangle_{u-d} &= a_{2,0}^{v}\left(1-2\left(\frac{M_{\pi}}{4\pi F_{0}}\right)^2\left((\overset{\circ}{g}_{A})^{2} + (3(\overset{\circ}{g}_{A})^{2} + 1)\log\left(\frac{M_{\pi}}{\mu}\right)\right)\right) + 4\frac{M_{\pi}^{2}}{m_{0}^{2}}c_{8}^{r}(\mu)\\ 
&\quad+ \frac{\overset{\circ}{g}_{A}M_{\pi}^{3}}{16\pi F_{0}^{2}m_{0}}\left(\frac{8}{3}\Delta a_{2,0}^{v} + \frac{7}{2}\overset{\circ}{g}_{A}a_{2,0}^{v} - \tilde{l}_{1}\right)\\ 
&\quad+ \left(\frac{M_{\pi}}{m_{0}}\right)^{4}\left(k_{1}\left(\log\left(\frac{M_{\pi}}{\mu}\right)\right)^{2} + k_{2}\log\left(\frac{M_{\pi}}{\mu}\right) + k_{3}\right) + \mathcal{O}(p^5).
\end{split}\label{eq:xumdexpansion}
\end{align}
For the LECs occuring in the leading-one-loop calculation we have used the nomenclature of \cite{Dorati:2007bk}, while at $\mathcal{O}(p^3)$, $\tilde{l}_{1}$ is a combination of the LECs $l_{1,n}$, $n\in\lbrace 6,7,15,16,18,19\rbrace$, defined in \cite{Wein:2014wma} (see Eq.~(A6c) in that reference). $m_{0},\overset{\circ}{g}_{A},F_{0}$ denote the nucleon mass, the axial coupling constant $g_{A}$ and the pion decay constant, respectively, in the two-flavor chiral limit. For the renormalization scale, we shall use $\mu = 1\,\mathrm{GeV}$ in the following. The terms of $\mathcal{O}(p^4)$, parameterized by the three constants $k_{i}$, are of two-loop order and have not been computed so far. Here we have made the reasonable assumption that the expansion of $\langle x\rangle_{u-d}$ at the two-loop level is analogous to the ones of the nucleon mass and $g_{A}$, compare \cite{McGovern:1998tm,Schindler:2006ha,Bernard:2006te}.\\In the following, we want to test the stability of the fits to lattice data employing the one-loop ChPT 
expressions for $\langle x\rangle_{u-d}$, with respect to variations of subleading (two-loop) order. To do this, we generate a large set of random numbers for $\lbrace k_{1},k_{2},k_{3}\rbrace$ and fit the three free one-loop parameters $a_{2,0}^{v},c_{8}^{r}(\mu=1\,\mathrm{GeV})$ and $\tilde{l}_{1}$ to data (see e.g. Sec.~7 of \cite{Bijnens:2011tb} for a similar strategy used to determine the Gasser-Leutwyler-LECs $L_{i}^{r}$). Of course, we should not let the coefficients $k_{i}$ become arbitrarily large. We shall assume that BChPT works reasonably well (along the lines of the usual low-energy power counting) for $M_{\pi}\lesssim\,200\,\mathrm{MeV}$. For Eq.~(\ref{eq:xumdexpansion}), this 
amounts to the constraint that $\left|k_{i}\right|\lesssim 2$. To get a robust estimate for possible higher-order effects, we will allow for a range $-4<k_{i}<+4$ for the random number sets.\\
We have yet to specify some input: For $\Delta a_{2,0}^{v}$, we take the same value as used in \cite{Dorati:2007bk}, which is consistent with information on $\langle x\rangle_{\Delta u-\Delta d}$ (compare e.g. \cite{Alexandrou:2014yha} for a recent overview). The value of $g_{A}$ in the chiral limit is not well known, and we simply substitute the phenomenological value here \cite{Beringer:1900zz}. The nucleon mass in the chiral limit is inferred from Tab.~B.4 of \cite{Bali:2012qs}, while the pion decay constant in this limit is taken from Tab.~1 of \cite{Baron:2009wt}. We collect 
the input values in Tab.~\ref{tab:input}.
\begin{table}[t]
\centering
\caption{Input values for the chiral fits.}
\label{tab:input}
\begin{ruledtabular}
\begin{tabular}{c c c c}
$m_{0}\,[\text{GeV}]$ & $\overset{\circ}{g}_{A}$ & $F_{0}\,[\text{GeV}]$ &  $\Delta a_{2,0}^{v}$  \\
\hline
$0.893$ & $1.270$ & $0.086$ & $0.210$ \\
\cite{Bali:2012qs} & \cite{Beringer:1900zz} & \cite{Baron:2009wt,Colangelo:2003hf} & \cite{Dorati:2007bk,Alexandrou:2014yha,Edwards:2006qx,Bratt:2010jn,Blumlein:2010rn}
\end{tabular}
\end{ruledtabular}
\end{table}
In a first step, we fit Eq.~(\ref{eq:xumdexpansion}) to recent lattice data published in \cite{Bali:2014gha} with $M_{\pi}<500\,\mathrm{MeV}$, selecting the largest available volumes. In the fits of type $1$, we exclude the point at $M_{\pi}\sim 150\,\mathrm{MeV}$ for now, which is measured at a rather small value of $M_{\pi}L\approx 3.5$, and is obviously inconsistent with the experimental value without applying finite volume (and possibly other) corrections. For a further discussion of this point, see Sec.~\ref{sec:disc}. The following Tab.~\ref{tab:fit1000} gives the results for fits of type $1$ with $k_{1,2,3}=0$.  In the fit marked with a prime, the last point at $M_{\pi}\sim 490\,\mathrm{MeV}$ has been dropped.\\
\begin{table}[h]
\centering
\caption{One-loop fit results ($k_{1}=k_{2}=k_{3}=0$) for the fits of type $1$.}
\begin{ruledtabular}
\begin{tabular}{c c c c c }
 fit & $a_{2,0}^{v}$ & $c_{8}^{r}$ & $\tilde{l}_{1}$ & $\chi^2/\mathrm{d.o.f.}$ \\
\hline
 $1$  & $0.134$ & $-0.184$ & $0.789$ & $0.182$\\
 $1'$ & $0.170$ & $-0.510$ & $0.190$ & $0.171$
\end{tabular} 
\end{ruledtabular}
\label{tab:fit1000}
\end{table}%
The cyan bands in the plots below consist of $10^3$ fits with prescribed random number values for the $k_{i}$. The black lines are the fit curves for $k_{i}=0$. 
\begin{figure}[h]
\centering
\subfigure[\,Fit scenario 1]{\includegraphics[width=0.44\textwidth]{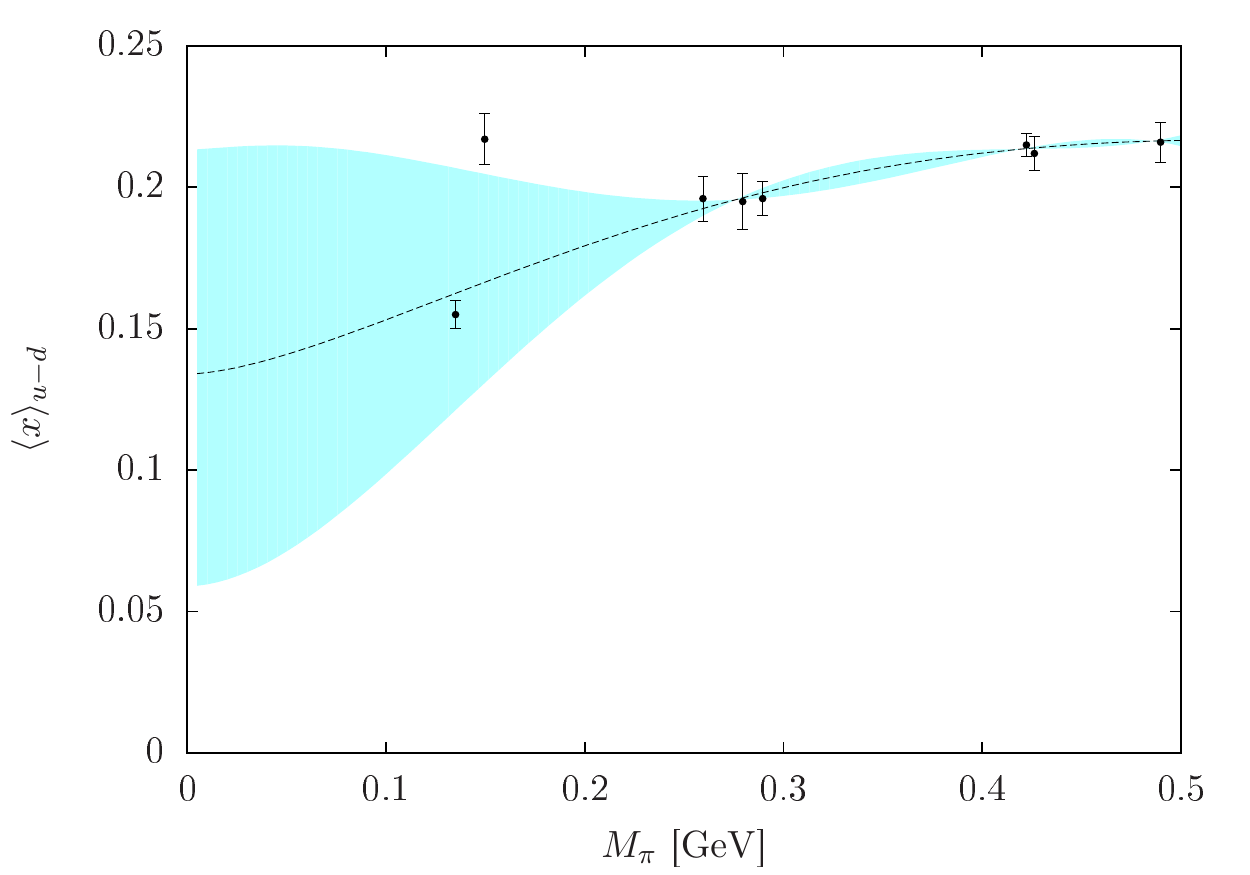}}\hspace{1cm}
\subfigure[\,Fit scenario 1']{\includegraphics[width=0.44\textwidth]{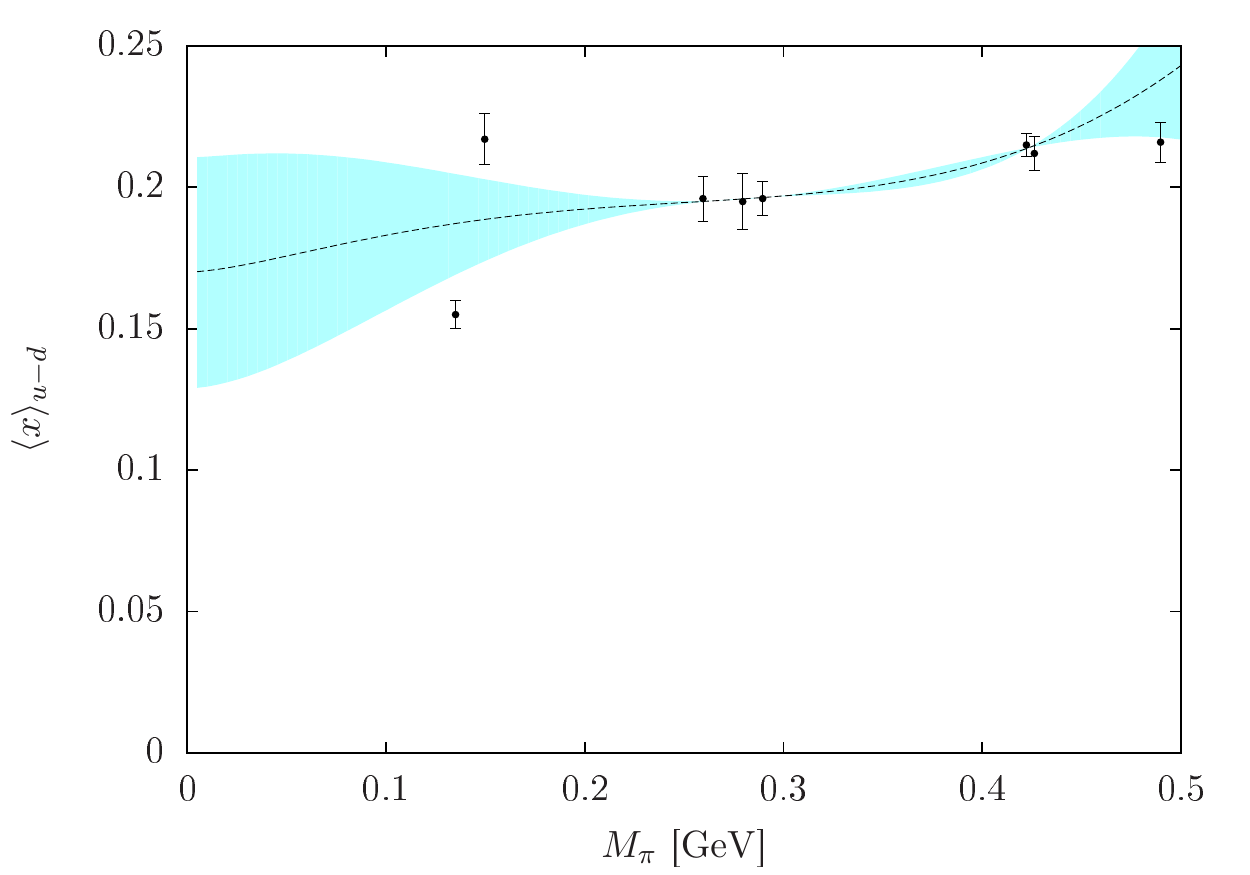}}
\caption{The results for fit scenarios 1 and 1'. The first point, which marks the experimental result, and the second point are not included in the fits.}
\label{fig:xumdbands1}
\end{figure}
The fact that the fit parameters and the resulting bands from fits 1 and 1' are quite different indicates that there is already a non-negligible sensitivity to uncontrolled higher-order effects for $M_{\pi}\approx 500\,\mathrm{MeV}$. But also note that there are only five available data points for three free parameters in fit 1'. Moreover, the convergence properties are problematic for this fit class, see below. \\

In the fits of type 2, we also include the phenomenological value taken as $\langle x\rangle_{u-d}^{\text{phen.}}=0.155\pm 0.005$ (following \cite{Green:2012ud}, see also \cite{Renner:2010ks} for a collection of values and references). However, we still exclude the point at $M_{\pi}\sim 150\,\mathrm{MeV}$. The following Tab.~\ref{tab:fit2000} displays the results for those fits with $k_{1,2,3}=0$.  In the fit marked with a prime, the last point at $M_{\pi}\sim 490\,\mathrm{MeV}$ has again been dropped.
\begin{table}[h]
\centering
\caption{One-loop fit results ($k_{1}=k_{2}=k_{3}=0$) for the fits of type $2$.}
\begin{ruledtabular}
\begin{tabular}{c c c c c }
 fit & $a_{2,0}^{v}$ & $c_{8}^{r}$ & $\tilde{l}_{1}$ & $\chi^2/\mathrm{d.o.f.}$ \\
\hline
$2$  & $0.125$ & $-0.116$ & $0.896$ & $0.196$ \\
$2'$ & $0.124$ & $-0.095$ & $0.947$ & $0.221$
\end{tabular} 
\end{ruledtabular}
\label{tab:fit2000}
\end{table}%
\begin{figure}[h]
\centering
\subfigure[\,Fit scenario 2]{\includegraphics[width=0.44\textwidth]{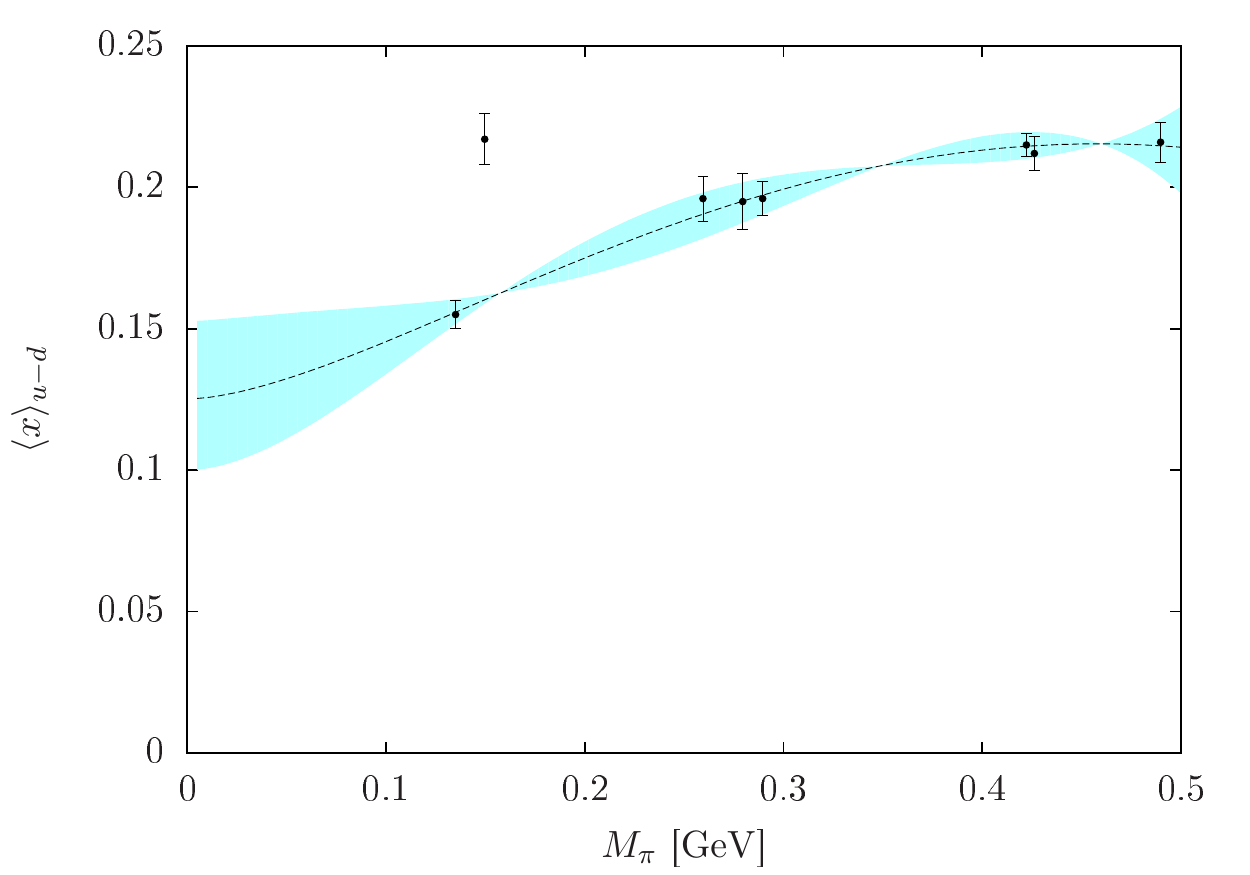}}\hspace{1cm}
\subfigure[\,Fit scenario 2']{\includegraphics[width=0.44\textwidth]{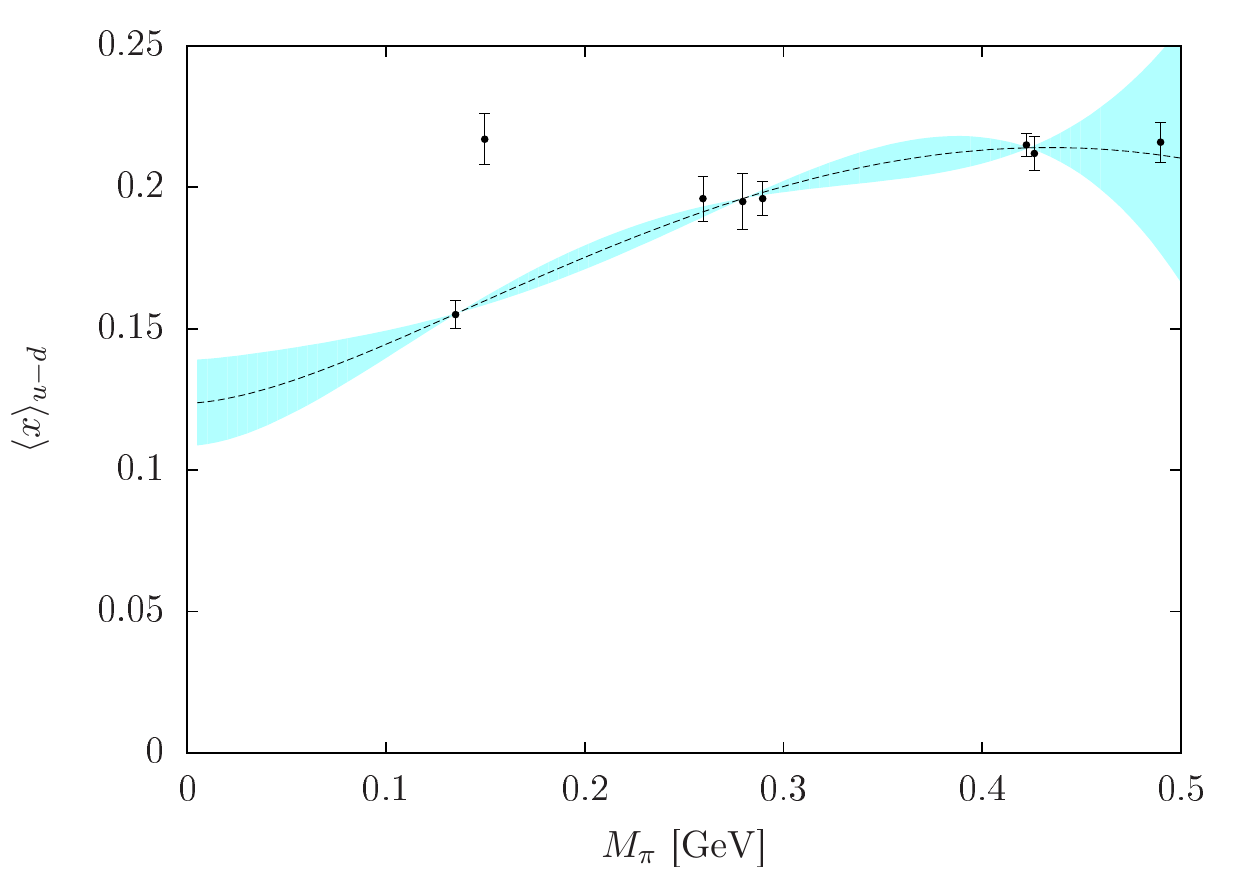}}
\caption{The results for fit scenarios 2 and 2'. Here, the experimental value (first point) is included in the fits.}
\label{fig:xumdbands2}
\end{figure}%
The inclusion of the experimental point greatly stabilizes the fits. It is striking that the value at $M_{\pi}\sim 150\,\mathrm{MeV}$ always lies outside the generated bands. Of course, finite volume effects might play a role here - see Sec.~\ref{sec:finite} for a discussion of this issue. Please also note that the fit curves lying close to the borders of the bands correpond to extrapolations with two-loop contributions which are quite large relative to the expectation from the chiral power counting, so the present way of quantifying the uncertainty in the extrapolation can be regarded as conservative in this respect. \\In an attempt to determine the LECs and their uncertainties, it makes sense to include the physical point. Collecting the resulting fit parameters for $10^4$ fits of type 2 results in the histograms in Figs.~\ref{fig:hist1}-\ref{fig:hist3}. The black curves are Gau{\ss} distributions fitted to the histograms, 
with the expectation values $\langle\mathrm{LEC}\rangle$ and $\sigma(\mathrm{LEC})$ as free parameters (given in Eq.~(\ref{eq:parsfit2}) below).
\begin{figure}[h]
\centering
\subfigure[\,Histogram for $a_{2,0}^v$]{\includegraphics[width=0.33\textwidth]{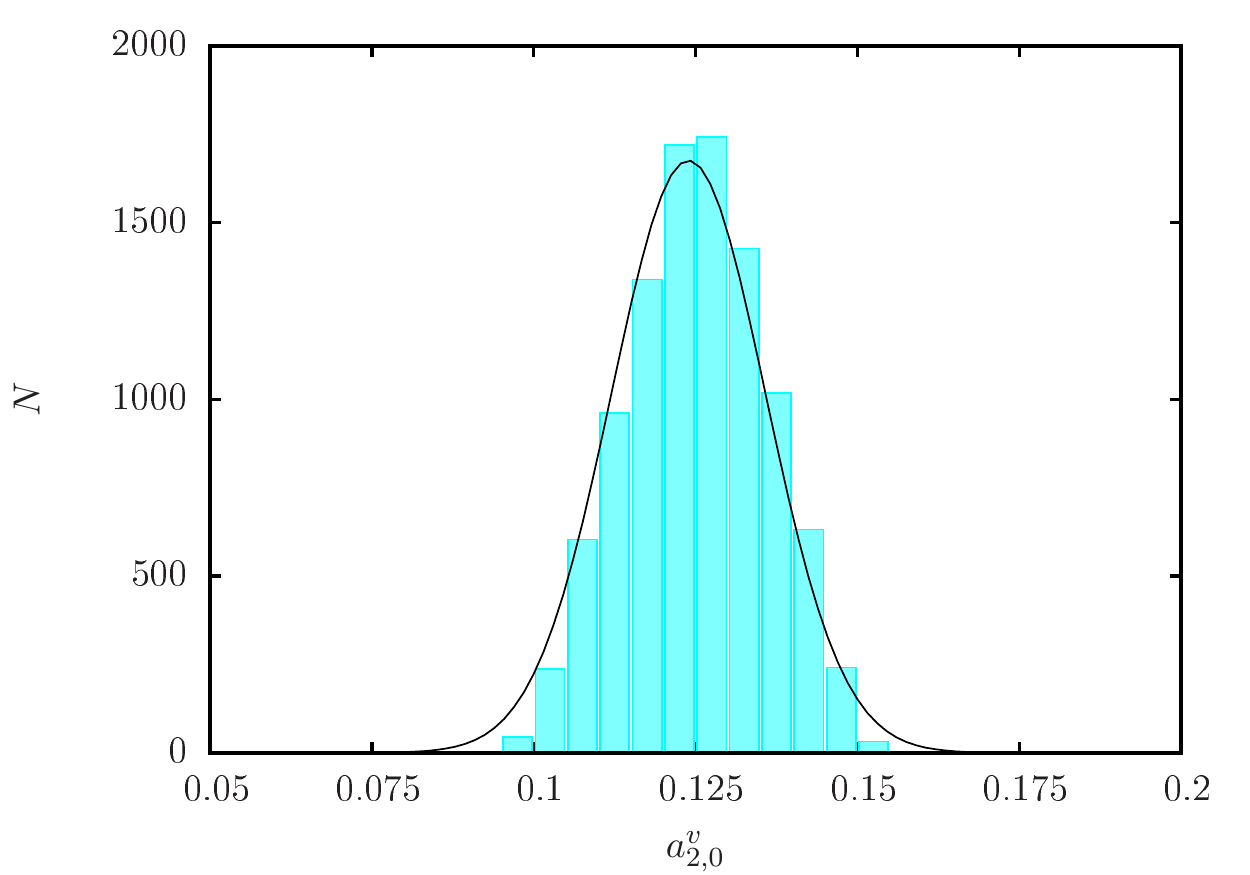}\label{fig:hist1}}
\subfigure[\,Histogram for $c_8^{(r)}$]{\includegraphics[width=0.33\textwidth]{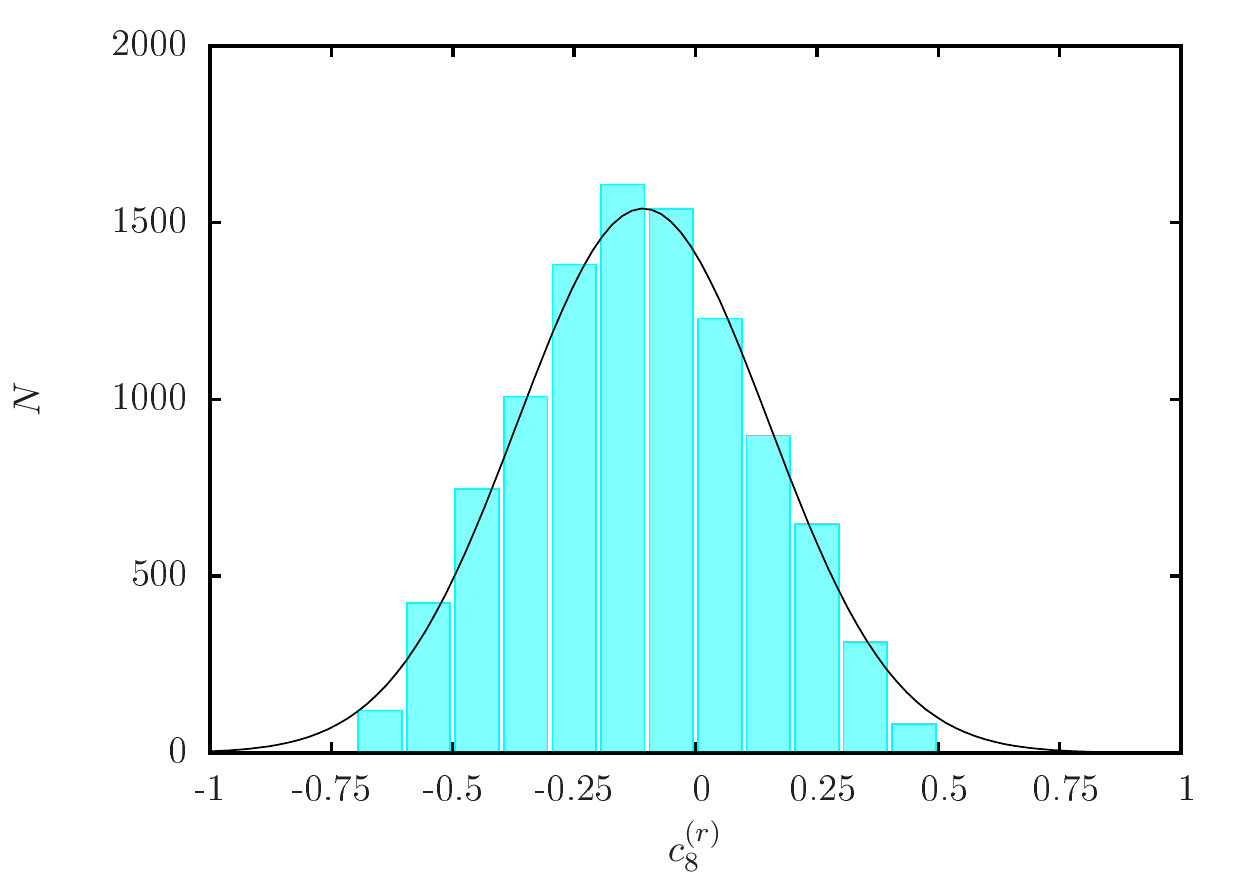}}
\subfigure[\,Histogram for $\tilde{l}_1$]{\includegraphics[width=0.33\textwidth]{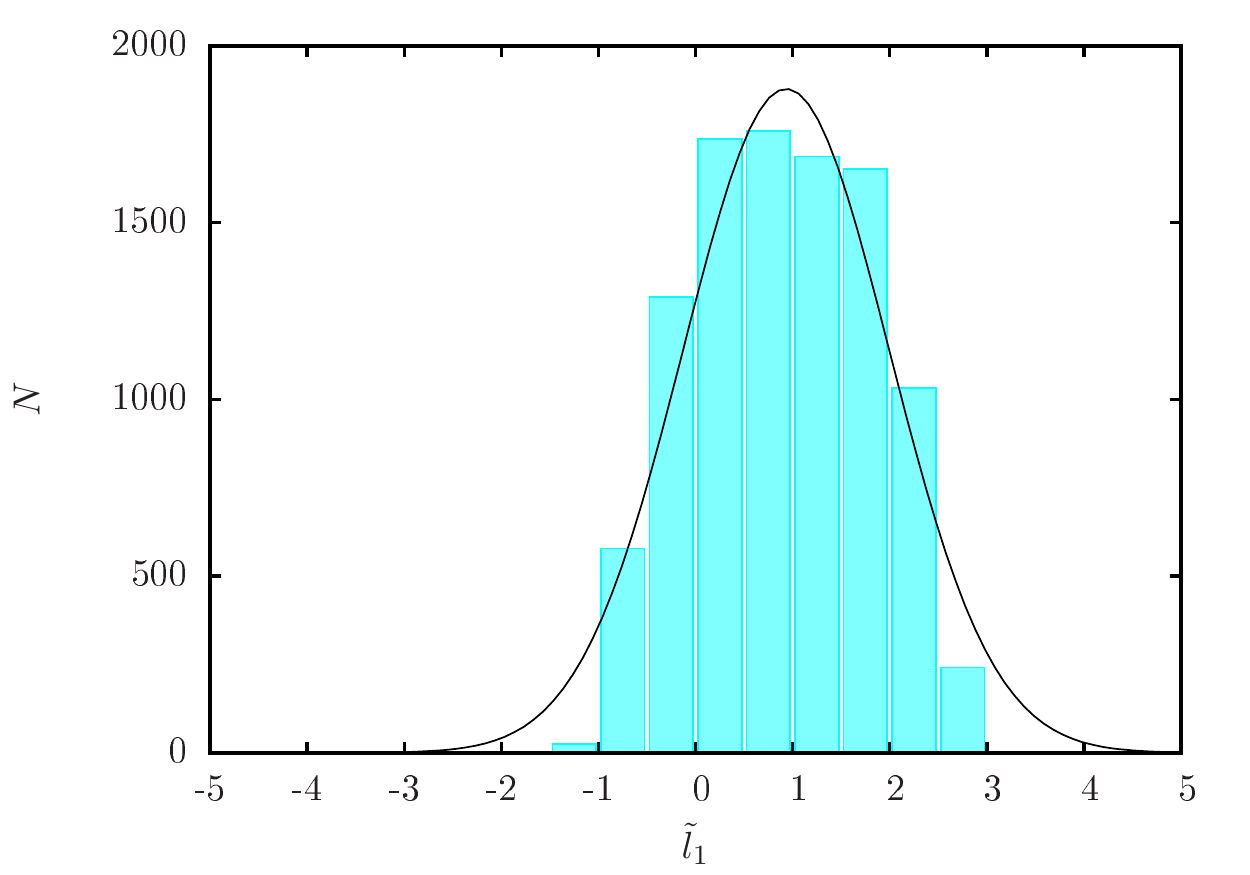}\label{fig:hist3}}
\caption{Histograms for $N=10^4$ fits including the fitted Gaussian distribution (black curve).}
\end{figure}
\begin{align}
\begin{split}
\langle a_{2,0}^{v}\rangle &= 0.125\,, \qquad \langle c_{8}^{r}(1\,\mathrm{GeV})\rangle = -0.115\,, \qquad\langle\tilde{l}_{1}\rangle = 0.899\,, \\
\sigma(a_{2,0}^{v}) &= 0.012\,, \qquad \sigma(c_{8}^{r}(1\,\mathrm{GeV})) =  0.256\,, \qquad \sigma(\tilde{l}_{1}) = 1.054\,. \label{eq:parsfit2}
\end{split}
\end{align}
We add some remarks on the convergence properties of the chiral expansion. For $M_{\pi}=200\,\mathrm{MeV}$ and the parameters from Tab.~\ref{tab:fit1000} and \ref{tab:fit2000},
\begin{eqnarray}
\langle x\rangle_{u-d} &=& a_{2,0}^{v}\left(1+\frac{\langle x\rangle_{u-d}^{(2)}}{a_{2,0}^{v}}+\frac{\langle x\rangle_{u-d}^{(3)}}{a_{2,0}^{v}}+\mathcal{O}(p^4)\right)\label{eq:conv}\\
 &\simeq& 0.125\left(1 + 0.345 + 0.054  + \ldots\right)\qquad\rm{for\,\, fit\, 2}\,,\nonumber\\
 &\simeq& 0.124\left(1 + 0.377 + 0.040  + \ldots\right)\qquad\rm{for\,\, fit\, 2'}\,,\nonumber\\
 &\simeq& 0.134\left(1 + 0.255 + 0.083  + \ldots\right)\qquad\rm{for\,\, fit\, 1}\,,\nonumber\\
 &\simeq& 0.170\left(1 - 0.071 + 0.202  + \ldots\right)\qquad\rm{for\,\, fit\, 1'}\nonumber\,.
\end{eqnarray}
Here the superscripts in round brackets denote the chiral order. It appears that the expansion converges well in the low-energy region except for the parameters from fit 1', for which no convergence is observed even at the physical point. On the one hand, this fit cannot be ruled out, since the fit parameters are still of natural size. But on the other hand, it is not very meaningful as already remarked above, because the included data points do not sufficiently constrain the three free parameters. In Fig.~\ref{fig:comp} we show plots for $\langle x\rangle_{u-d}^{(2)}/a_{2,0}^{v}$ (red) and  $\langle x\rangle_{u-d}^{(3)}/a_{2,0}^{v}$ (black, dotted) for fit 2 and 2'. It seems that the application of the one-loop approximation becomes problematic somewhere between $M_{\pi}\sim 300\ldots 450\,\mathrm{MeV}$. This is in accord with our findings in Sec.~VII of \cite{Bruns:2012eh} for the case of baryon masses, and with the general expectation stated e.g. in \cite{Bernard:2007zu}. The plotted curves also demonstrate that it is illegitimate to neglect the $M_{\pi}^3$-term for data with $M_{\pi}\gtrsim 300\,\mathrm{MeV}$, as was often done in applications. Obviously, some more data points with $M_{\pi}\lesssim\,350\,\mathrm{MeV}$ and bigger volumes are necessary to bring the extrapolation under better control.\\
\begin{figure}[h]
\centering
\subfigure[\,Fit scenario 2]{\includegraphics[width=0.44\textwidth]{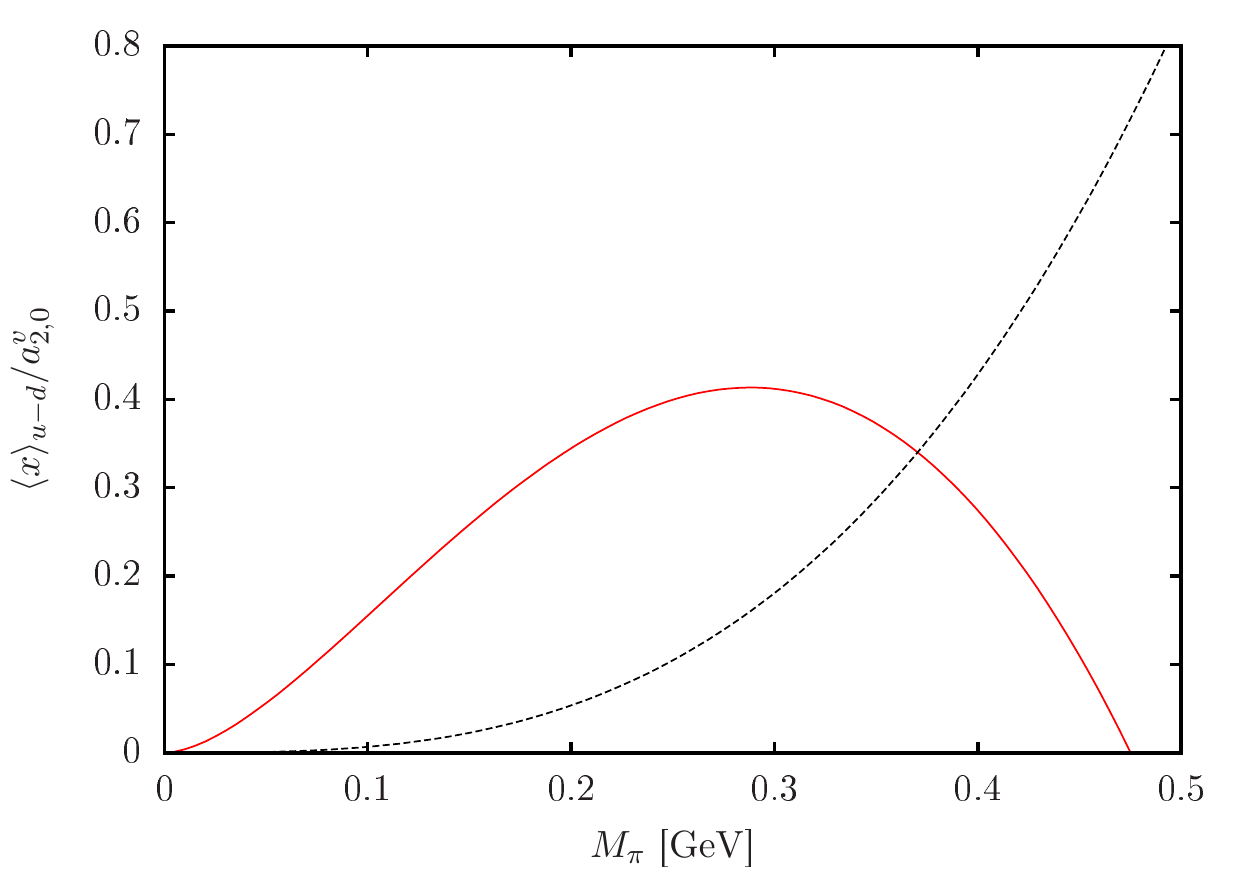}}\hspace{1cm}
\subfigure[\,Fit scenario 2']{\includegraphics[width=0.44\textwidth]{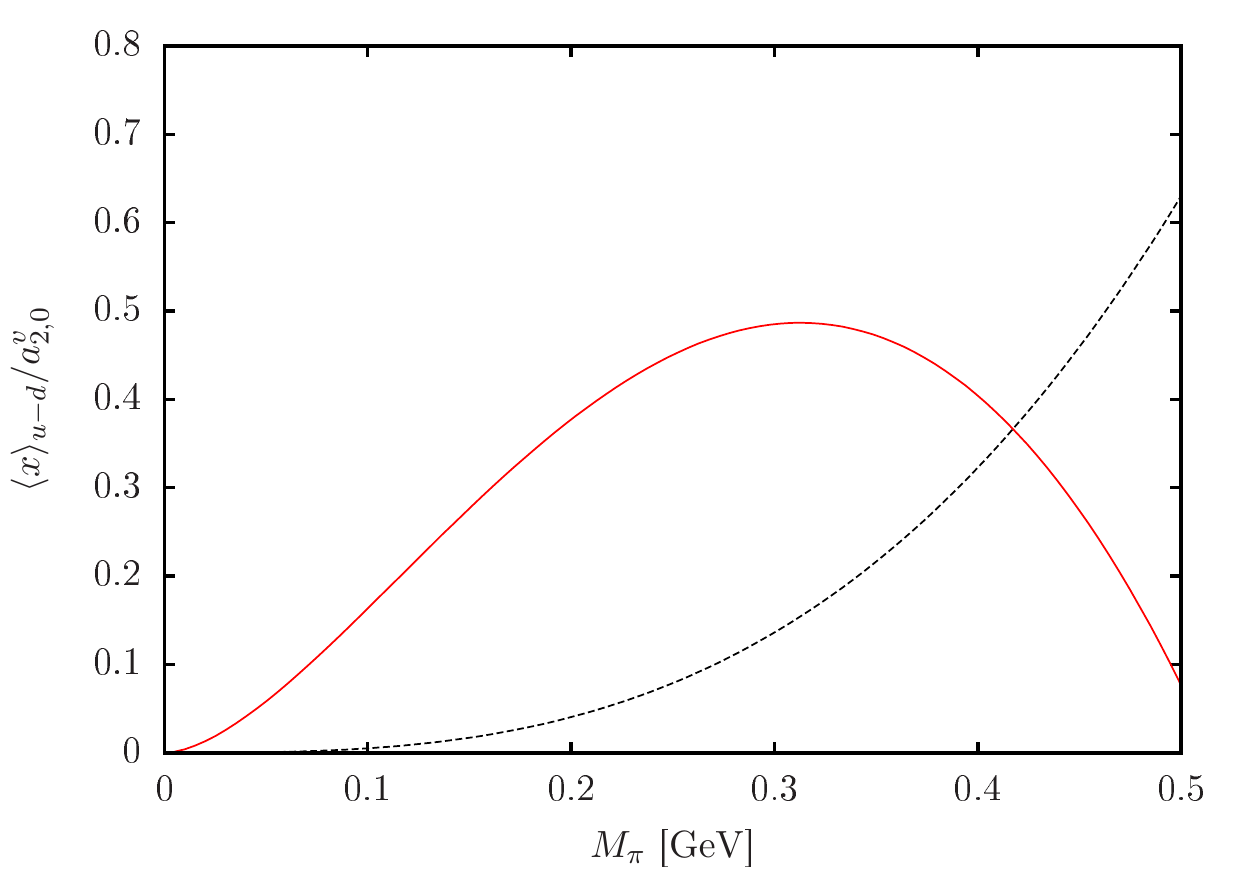}}
\caption{Comparison of second order (red) to third order (black) contribution to $\langle x\rangle_{u-d}/a_{2,0}^v$.}
\label{fig:comp}
\end{figure}%
As a further experiment, we include the lattice point at $M_{\pi}\sim 150\,\mathrm{MeV}$ as it stands, instead of the experimental input, neglecting possible finite volume effects for the moment. Let us call this fit scenario 3. As we can see, the $\chi^2/\mathrm{d.o.f.}$ becomes worse roughly by a factor of twelve compared to the previous fits, though there are some rare fits of similar quality (necessitating large higher-order effects), as can be read off from the histogram of $\chi^2$-values in Fig.~\ref{fig:chi2}. Also, the experimental point is always far outside the uncertainty band given by the estimate of the two-loop effects.
\begin{table}[h]
\centering
\caption{One-loop fit results ($k_{1}=k_{2}=k_{3}=0$)}
\begin{ruledtabular}
\begin{tabular}{c c c c c }
 fit & $a_{2,0}^{v}$ & $c_{8}^{r}$ & $\tilde{l}_{1}$ & $\chi^2/\mathrm{d.o.f.}$ \\
\hline
 $3$  & $0.183$ & $-0.562$ & $0.196$ & $2.571$
\end{tabular} 
\end{ruledtabular}
\label{tab:fit3}
\end{table}%
\begin{figure}[h]
\centering
\subfigure[\,Fit scenario 3]{\includegraphics[width=0.44\textwidth]{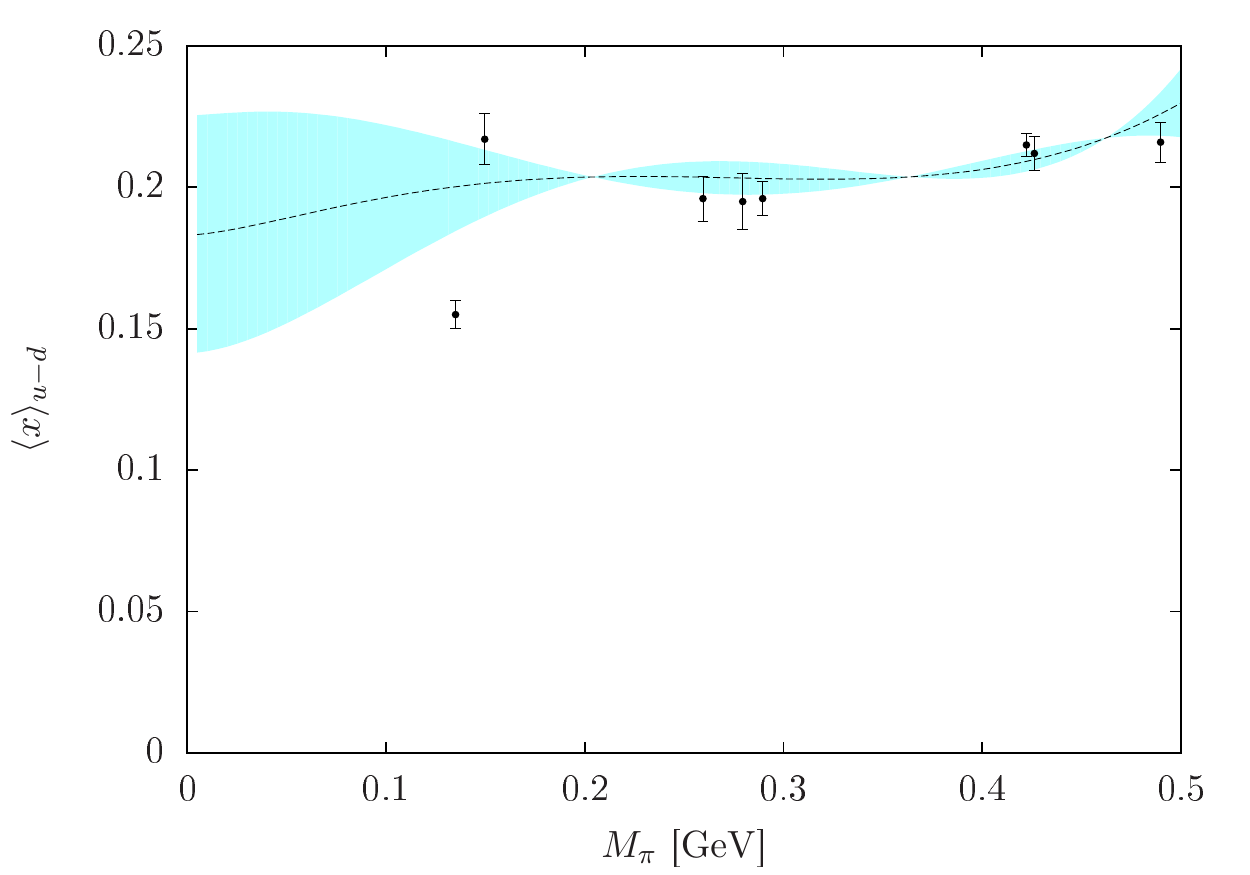}}\hspace{1cm}
\subfigure[\,Histogram for reduced $\chi^2$]{\includegraphics[width=0.44\textwidth]{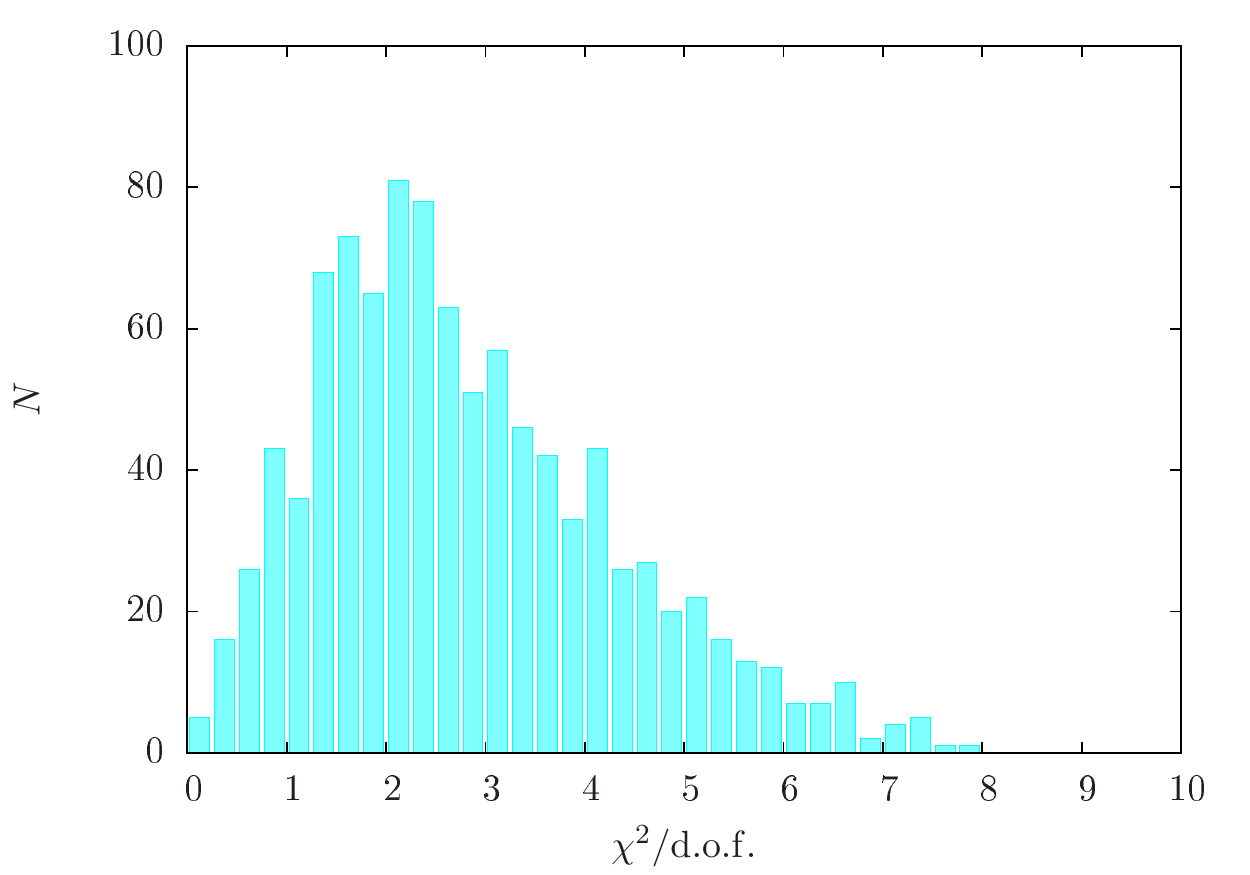}\label{fig:chi2}}
\caption{Results for fit scenario 3, where the lowest lattice data point is included.}
\label{fig:fit3}
\end{figure}%
No sign of convergence at $M_{\pi}=200\,\mathrm{MeV}$ is visible for this fit class,
\begin{align}
\langle x\rangle_{u-d} \simeq 0.183\left(1 - 0.085 + 0.196  + \ldots\right)\qquad\rm{for\,\, fit\, 3}\,.
\end{align}
This pattern is similar to that for fit $1'$. Including some more data points from other collaborations (the LHPC point of \cite{Bratt:2010jn} in the large volume (red) and the two RBC points of \cite{Aoki:2010xg} for $M_{\pi}<500\,\mathrm{MeV}$ (orange)) does not change the general picture established above, see Figs.~\ref{fig:RBC1}-\ref{fig:RBC2}. The $\chi^2/\mathrm{d.o.f.}$ values for the corresponding fits lie between $0.7\ldots 2.0$. The dashed black lines correspond to the previous fits from Tab.~\ref{tab:fit1000} and \ref{tab:fit2000}. Recall that the experimental value is not included in fit $1$ and $1'$. The pertaining parameter sets are given in Tab.~\ref{tab:fit000RBC}. The main difference to the previous fits is the fact that the 'flat' behavior of fit $1'$ is now altered to a curve that strongly bends down in the chiral regime, similar to the other curves, which is mainly due to the high-statistics LHPC point at $M_{\pi}\approx 356\,\mathrm{MeV}$, $M_{\pi}L\approx 6.3$. We shall see in the next section that this instability of the first fit scenario (the different behavior of fits $1$ and $1'$) is eliminated when the additional information on the modifications in a finite volume is taken into account.
\begin{figure}[h]
\centering
\subfigure[\,Fit scenario 1]{\includegraphics[width=0.44\textwidth]{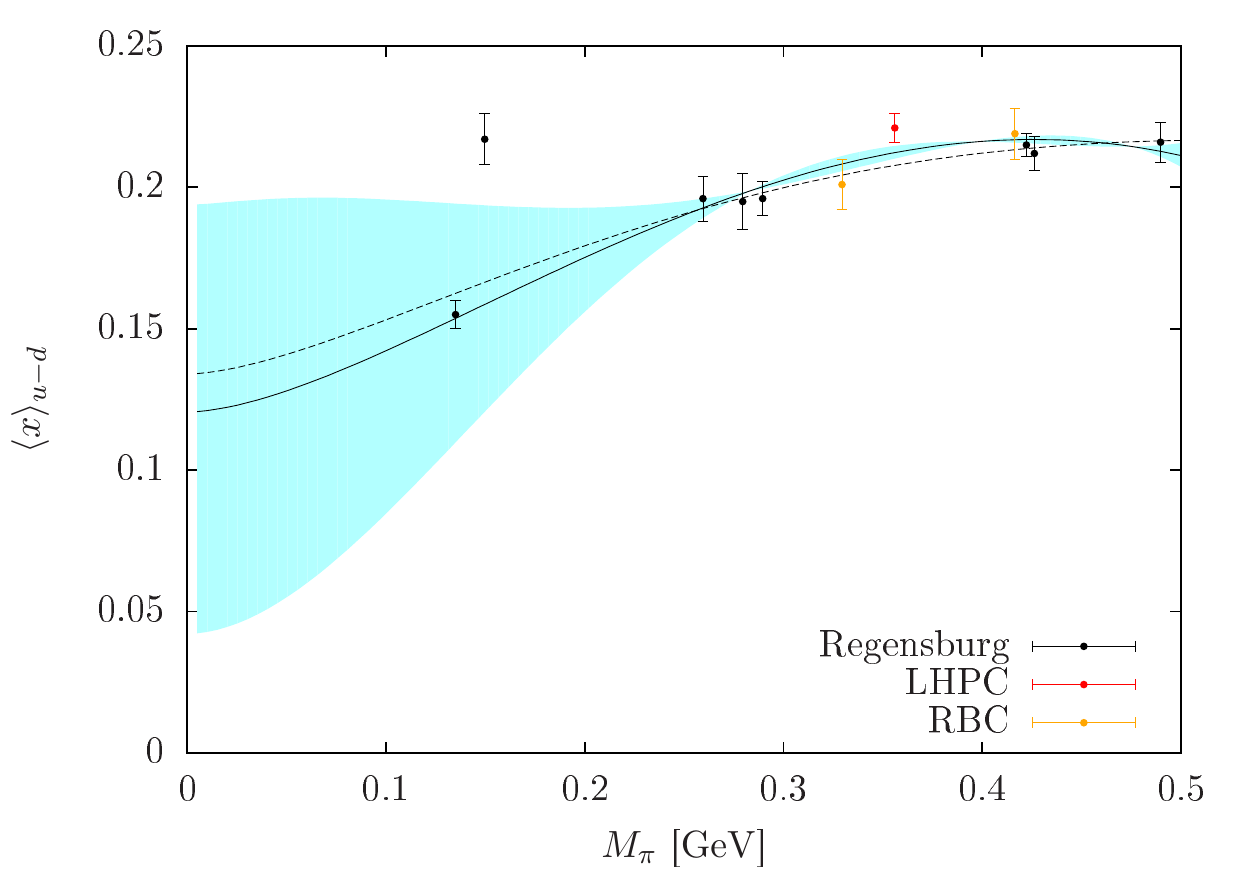}\label{fig:RBC1}}
\subfigure[\,Fit scenario 1']{\includegraphics[width=0.44\textwidth]{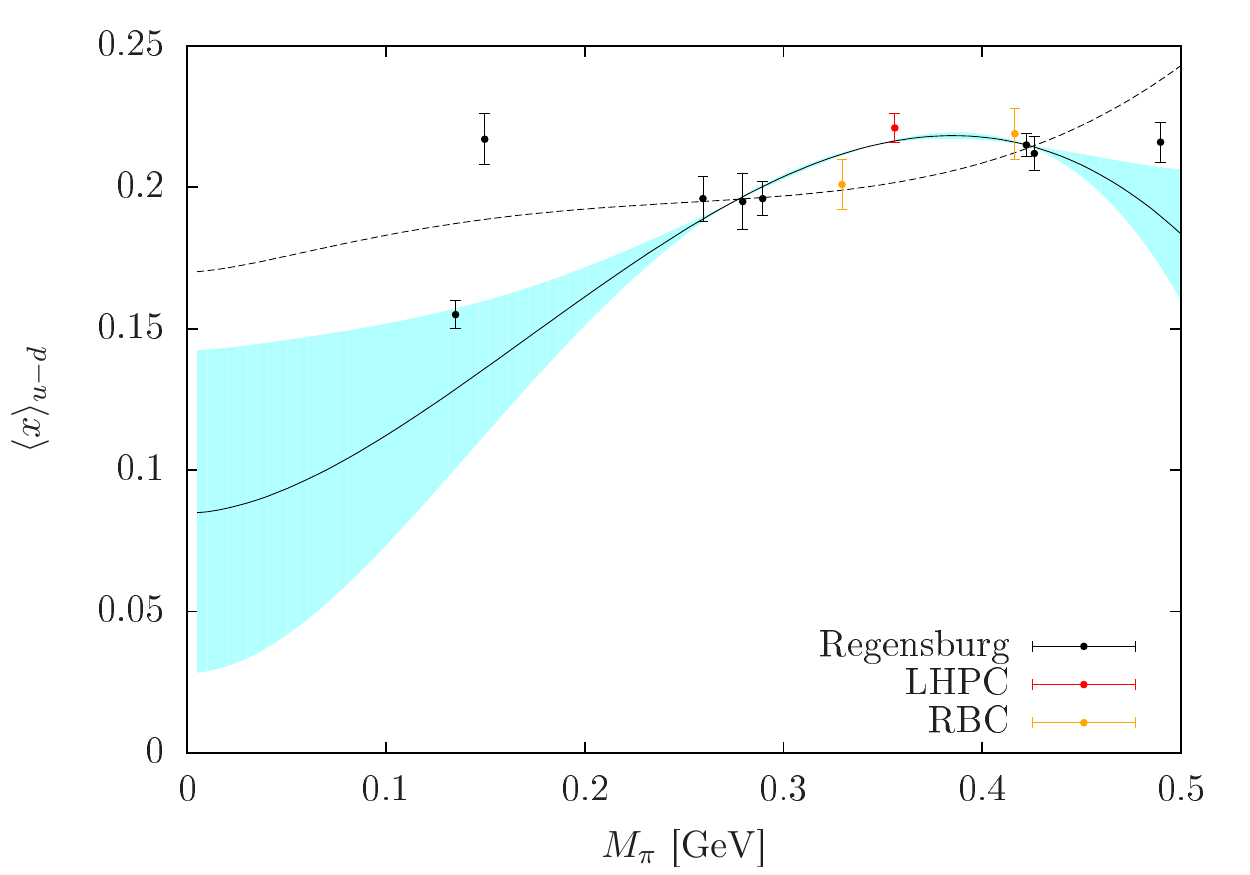}\label{fig:RBC1prime}}
\subfigure[\,Fit scenario 2]{\includegraphics[width=0.44\textwidth]{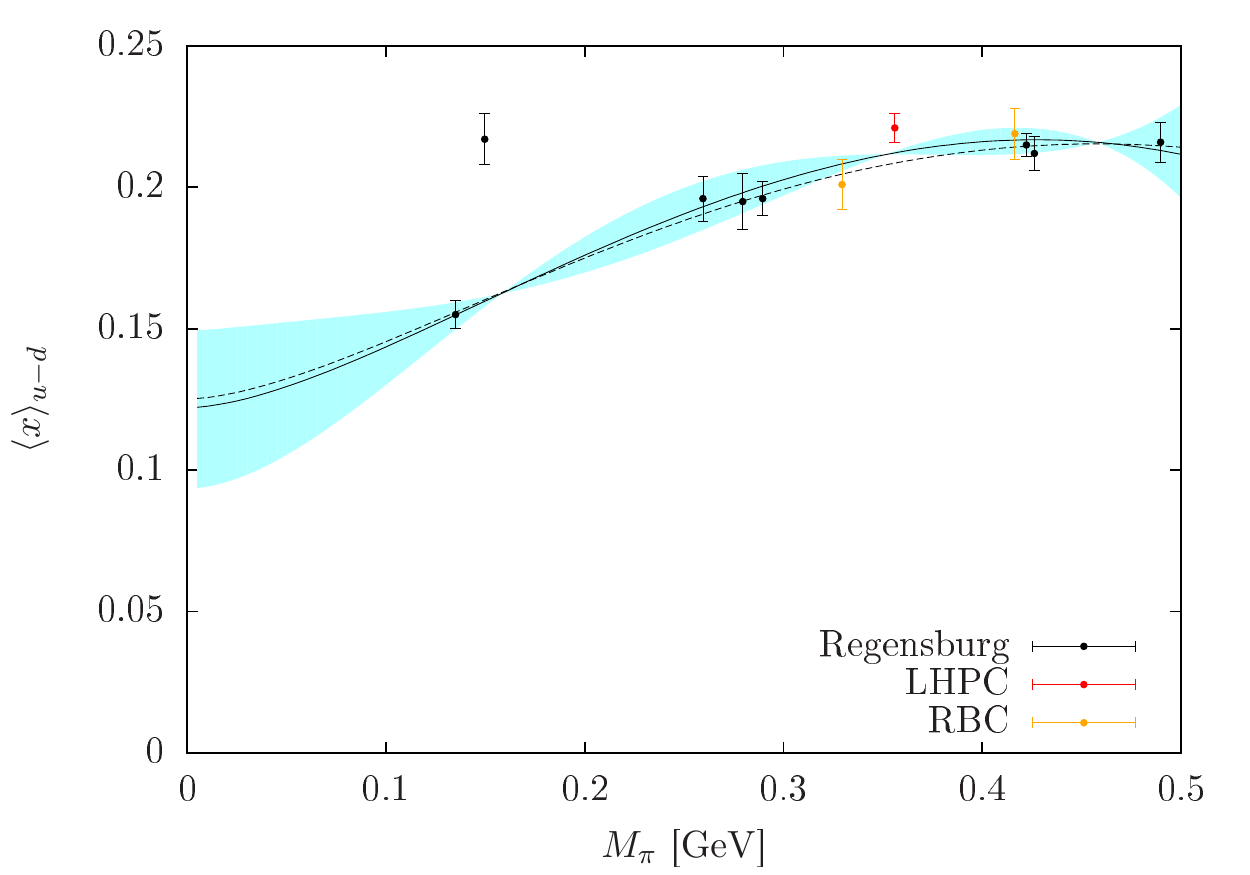}\label{fig:RBC2}}
\caption{Results for fit scenario 1, 1' and 2 with added data points from RBC and LHPC groups. The dashed line corresponds to the previous results without the added data points.}
\end{figure}
\begin{table}[htp]
\caption{One-loop fit results with additional data points ($k_{1}=k_{2}=k_{3}=0$)}
\begin{minipage}{0.45\textwidth}
\begin{ruledtabular}
\begin{tabular}{c c c c c}
 fit & $a_{2,0}^{v}$ & $c_{8}^{r}$ & $\tilde{l}_{1}$ & $\chi^2/\mathrm{d.o.f.}$ \\
\hline
 $1$  & $0.121$ & $-0.060$ & $1.011$ & $0.934$\\
 $1'$ & $0.085$ & $+0.256$ & $1.589$ & $0.773$
\end{tabular}
\end{ruledtabular}
\end{minipage}\hspace{1cm}
\begin{minipage}{0.45\textwidth}
\begin{ruledtabular}
\begin{tabular}{c c c c c}
fit & $a_{2,0}^{v}$ & $c_{8}^{r}$ & $\tilde{l}_{1}$ & $\chi^2/\mathrm{d.o.f.}$  \\
\hline
 $2$  & $0.122$ & $-0.071$ & $0.994$ & $0.802$ \\
 $2'$ & $0.120$ & $-0.036$ & $1.078$ & $0.846$
\end{tabular}
\end{ruledtabular}
\end{minipage}
\label{tab:fit000RBC}
\end{table}
Concerning the fits of type $3$, we also observe that the picture remains qualitatively the same as given in Tab.~\ref{tab:fit3} above, when including additional data points. The corresponding results are given in Tab.~\ref{tab:fit3RBC} below. Here we have again added a fit where the data point with the highest pion mass has been excluded from the $\chi^2$ function (labeled as fit $3'$). The pertaining curves look almost exactly the same as in Fig.~\ref{fig:fit3}. Again we find that the $\chi^2/\mathrm{d.o.f.}$ values are significantly larger than in the fit scenarios $1,2$, in which the point at $M_{\pi}\sim 150\,\mathrm{MeV}$ is {\em not}\, included. 
\begin{table}[h]
\centering
\caption{One-loop fit results with additional data points, including the lowest lattice data point.}
\begin{ruledtabular}
\begin{tabular}{c c c c c }
 fit & $a_{2,0}^{v}$ & $c_{8}^{r}$ & $\tilde{l}_{1}$ & $\chi^2/\mathrm{d.o.f.}$ \\
\hline
 $3$   & $0.174$ & $-0.462$ & $0.392$ & $2.784$\\
 $3'$  & $0.189$ & $-0.614$ & $0.071$ & $2.669$
\end{tabular} 
\end{ruledtabular}
\label{tab:fit3RBC}
\end{table}%
\newpage
\section{Finite volume analysis\label{sec:finite}}
As was mentioned in the foregoing section, the data point at $M_{\pi}\approx 150\,\text{MeV}$ always lies outside of the generated stability bands, which could be due to several systematic errors inherent to every LQCD simulation. In this section we want to analyze how large the finite volume corrections to $\langle x\rangle_{u-d}$ are and whether they can explain the discrepancies that were found in \cite{Bali:2014gha} (see also \cite{Alexandrou:2013jsa} for a consistent measurement with somewhat larger error bars). We utilize the full one-loop finite volume corrections calculated in \cite{Greil:2014awa} and the input parameters from Tab.~\ref{tab:input}. The LECs $l_{1,n}$ used in \cite{Wein:2014wma,Greil:2014awa} are set to zero for $n=\{1,3,6,7,8,13,14,15,16\}$ and the combination $(l_{1,18}+l_{1,19})$ is treated as a free fit parameter we call $l_{1,18+19}$. This is an allowed prescription at the order we are working, because there is only one free parameter in the $M_{\pi}^3$-term of $\langle x\rangle_{u-d}$, compare Eq.~(\ref{eq:xumdexpansion}).
To the same order $p^3$ in the chiral counting, we then find that this parameter can be related to $\tilde{l}_1$ from the previous section as follows:
\begin{align}
l_{1,18+19} = \frac{1}{32m_0}\left(\tilde{l}_1-\frac{4}{3}\Delta a^v_{2,0}\right).
\end{align}
For the finite volume analysis, and the comparison with the results of the previous section, we focus on the data set published in \cite{Bali:2014gha}. In this work, particular care has been taken to discriminate the ground state signal from excited state contributions, which is known to be very relevant for $\langle x\rangle_{u-d}$ \cite{Green:2012ud,Dinter:2011sg,Green:2011fg,Bali:2012av,Jager:2013kha}.
Just like in the previous section, we explore three different scenarios: in fit scenario $1\,\mathrm{fv}$ we fit all data with $200\,\text{MeV}<M_{\pi}<500\,\text{MeV}$ and in fit scenario $2\,\mathrm{fv}$ we include the experimental point at $M_{\pi}\approx135\,\text{MeV}$. In fit scenario $3\,\mathrm{fv}$ we fit the lattice data including the point at $M_{\pi}\approx150\,\text{MeV}$. The fits marked with a prime are variants where the last data point at $M_{\pi}\sim 490\,\mathrm{MeV}$ is excluded. We add the three points from the RBC and LHPC collaborations \cite{Bratt:2010jn,Aoki:2010xg} also used for the fits of Figs.~\ref{fig:RBC1}-\ref{fig:RBC2}.\\
\newpage
\begin{figure}[!h]
\centering
\subfigure[\,Fit scenario 1fv]{\includegraphics[width=0.45\textwidth]{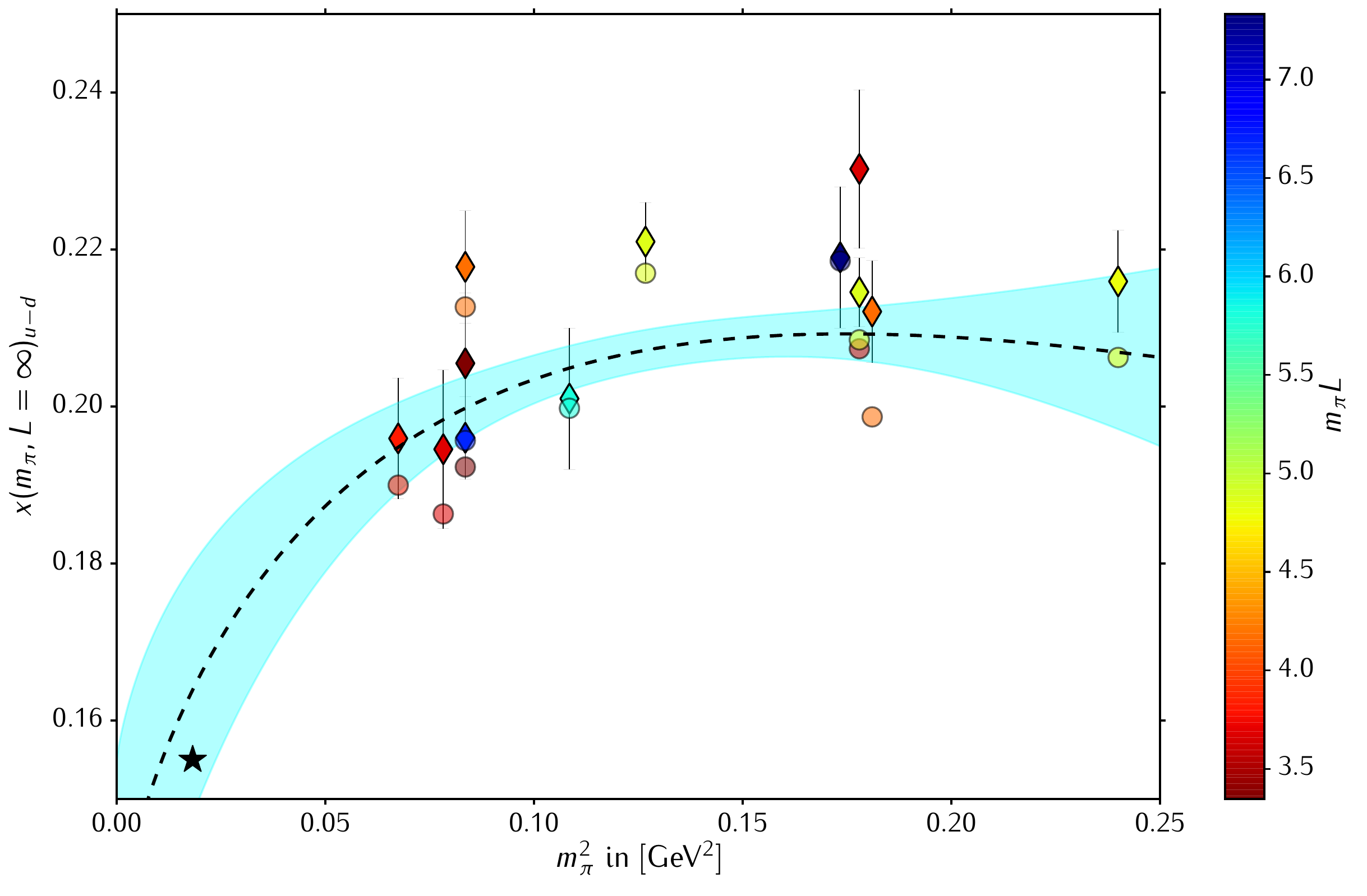}}
\subfigure[\,Fit scenario 1'fv]{\includegraphics[width=0.45\textwidth]{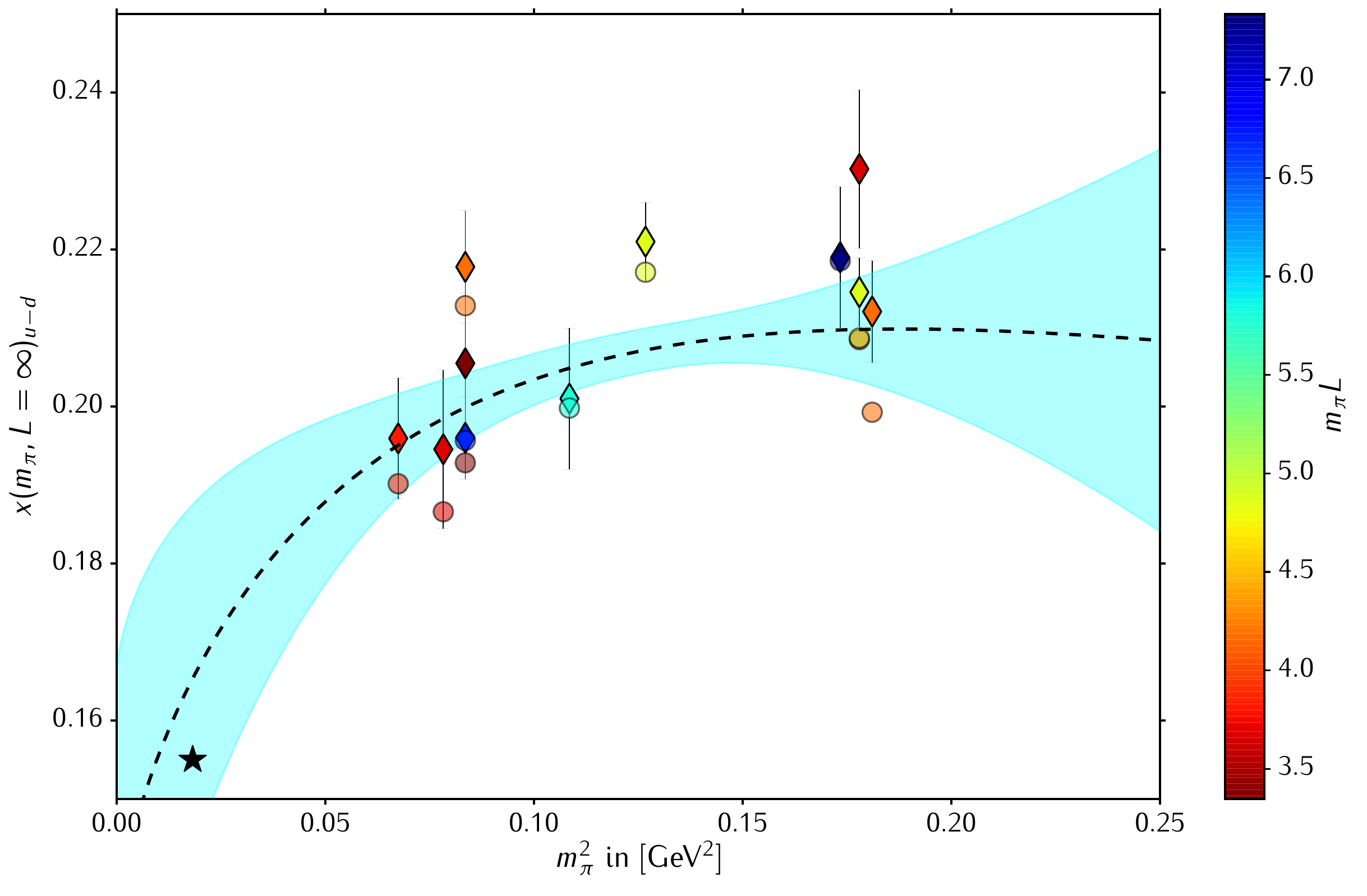}}\\
\subfigure[\,Fit scenario 2fv]{\includegraphics[width=0.45\textwidth]{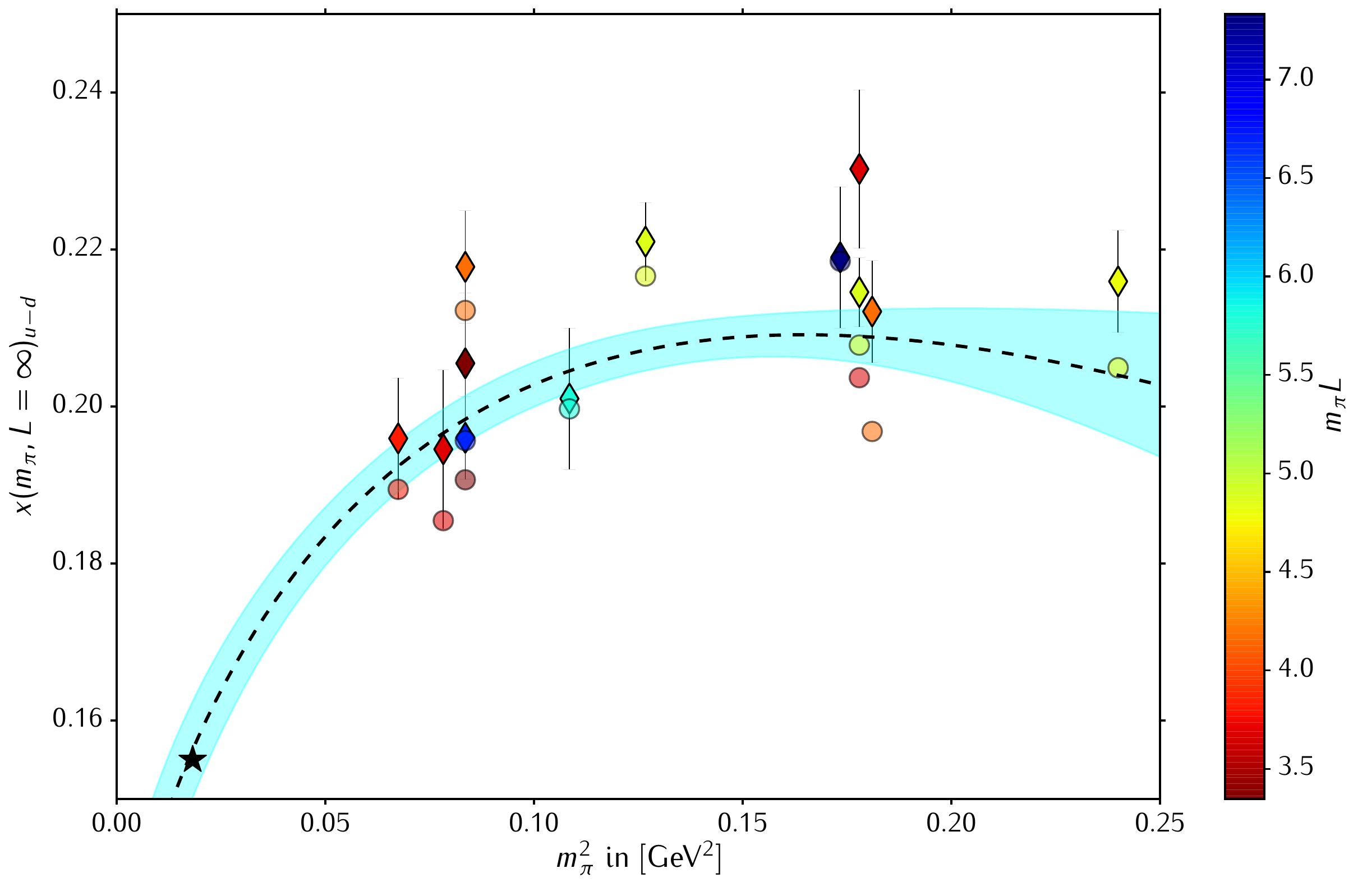}}
\subfigure[\,Fit scenario 2'fv]{\includegraphics[width=0.45\textwidth]{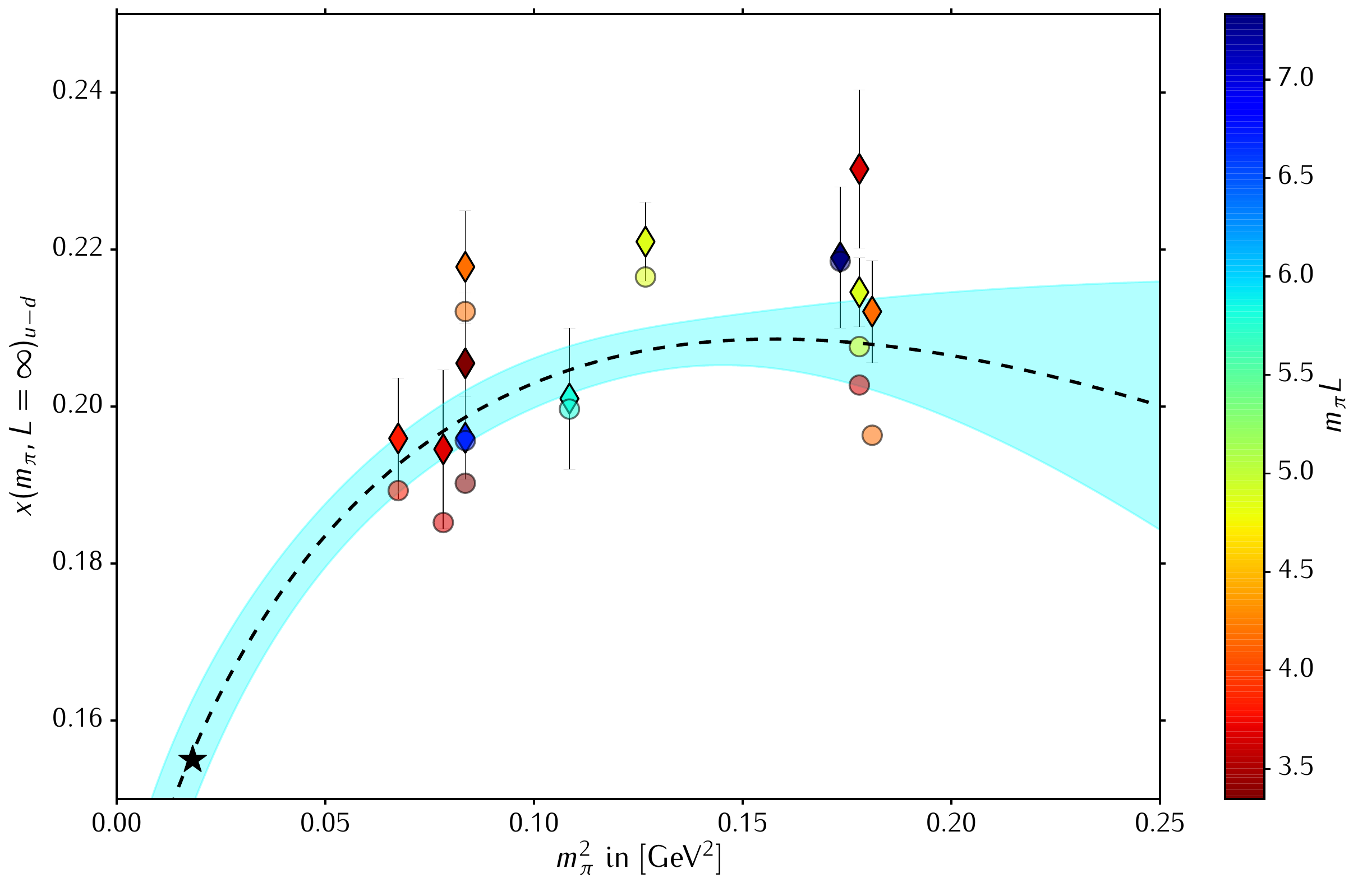}}\\
\subfigure[\,Fit scenario 3fv]{\includegraphics[width=0.45\textwidth]{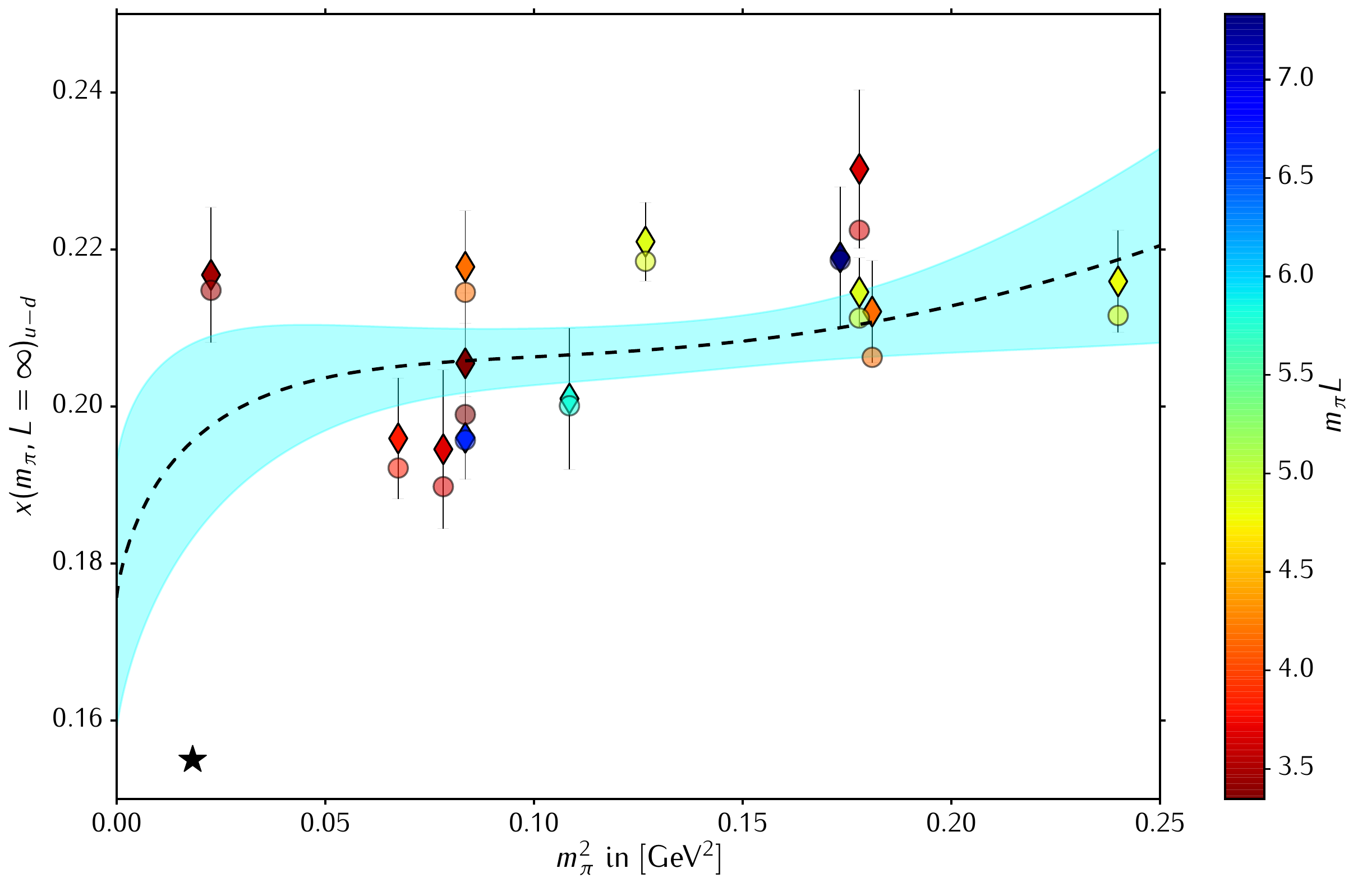}}
\subfigure[\,Fit scenario 3'fv]{\includegraphics[width=0.45\textwidth]{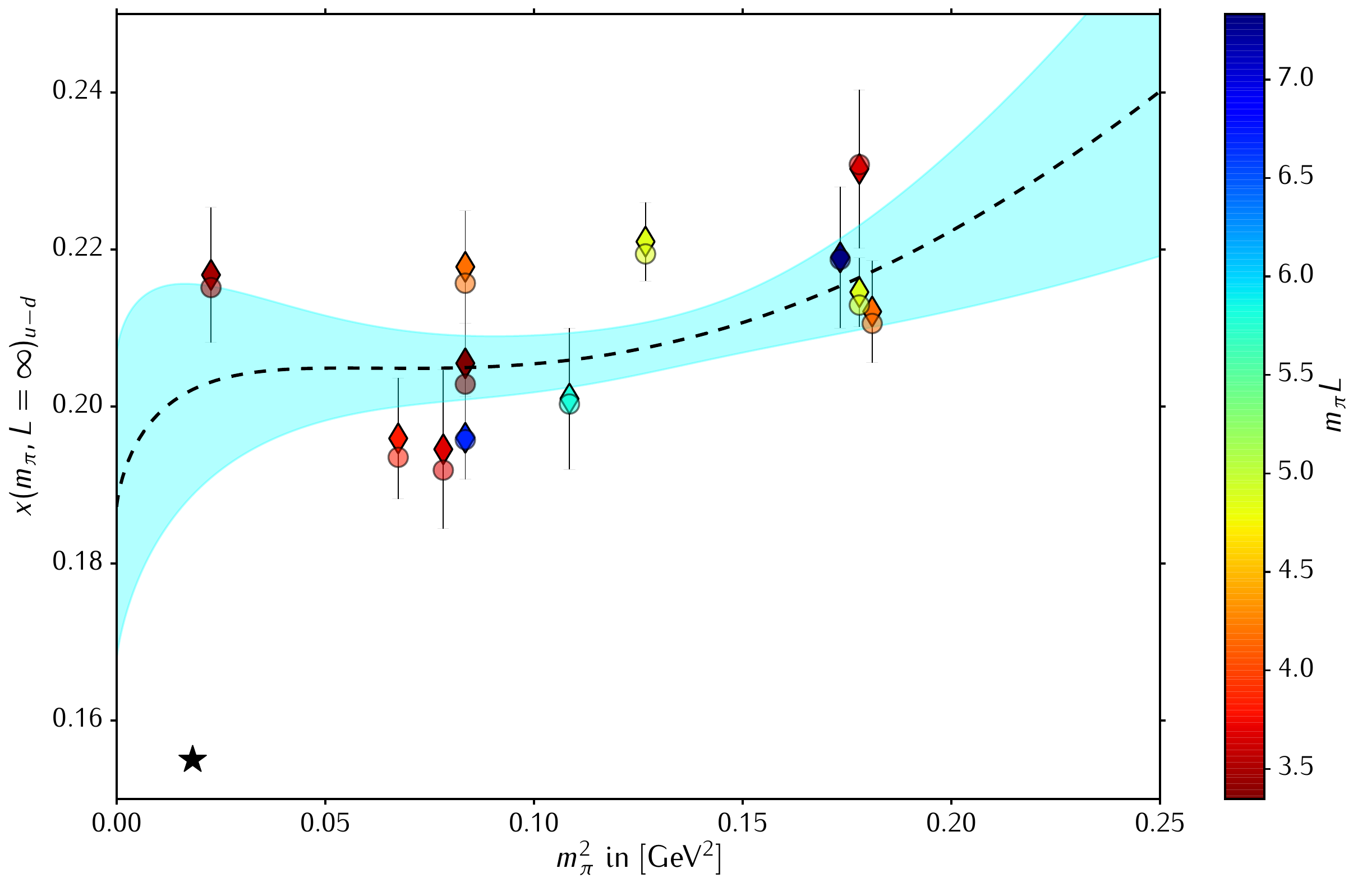}}
\caption{The figure shows the results for all fit scenarios. The finite volume corrected data points are represented by circles, while the raw data is represented by diamonds, and the experimental result is marked with a star. The shaded band represents the one sigma error band.}
\label{fig:fvfits}
\end{figure}%
\newpage
\begin{table}
\centering
\caption{Fit results including finite volume corrections for all fit scenarios.}
\label{tab:fvfits}
\begin{ruledtabular}
\begin{tabular}{c c c c c c}
scenario & $a_{2,0}^v$ & $c^r_8(1\,\text{GeV})$ & $l_{1,18+19}\,[\text{GeV}^{-1}]$ & $\tilde{l}_1$  & $\chi^2/\text{d.o.f.}$\\
\hline
$1$ fv   & $0.133(17)$ & $-0.136(147)$ & $0.026(15)$  & $1.024(702)$  & $1.53$\\
$1'$ fv  & $0.135(23)$ & $-0.159(217)$ & $0.024(23)$  & $0.952(938)$  & $1.72$\\
$2$ fv   & $0.123(7)$  & $-0.048(65)$  & $0.035(8)$   & $1.271(488)$  & $1.58$\\
$2'$ fv  & $0.122(7)$  & $-0.034(79)$  & $0.037(6)$   & $1.324(553)$  & $1.77$\\
$3$ fv   & $0.176(11)$ &  $-0.504(96)$ & $-0.010(10)$ & $0.005(568)$  & $2.47$\\
$3'$ fv  & $0.187(10)$ & $-0.645(87)$  & $-0.027(11)$ & $-0.490(627)$ & $2.35$
\end{tabular}
\end{ruledtabular}
\end{table}%
We find that, for scenarios $3\,\mathrm{fv}$ and $3'\,\mathrm{fv}$, the LECs $a_{2,0}^v$ and $c^r_8$ lie outside of the bounds generated in Sec.~\ref{sec:stab} and that these fit results have significant overlap with the results of fit $3$ from Sec.~\ref{sec:stab}. All in all, our fit results for scenarios $1\,\mathrm{fv}$, $2\,\mathrm{fv}$,  $2'\,\mathrm{fv}$ and noteably also $1'\mathrm{fv}$ are compatible with the bounds obtained from our stability considerations presented in the previous section, which is a very encouraging result.\\ The results above show that finite volume corrections are too small to account for the discrepancy found between the data point at $M_{\pi}\approx150\,\text{MeV}$ and the well-known experimental value at $M_{\pi}\approx135\,\text{MeV}$.\\ As was mentioned in the introduction, all finite volume formulae obtained from ChPT in the $p$-expansion are valid for $M_{\pi}L\gg1$ \cite{Colangelo:2006mp,Colangelo:2003hf}. It stands to debate whether $M_{\pi}L\approx3.5$ is already large enough for a legitimate application of these $p$-expanded finite volume 
formulae. Investigating this question would exceed the limits set for this analysis though. Note that no significant finite volume effects in the data were observed in \cite{Bali:2014gha}, in agreement with the findings of \cite{Bratt:2010jn,Aoki:2010xg,Green:2012ud,Alexandrou:2011nr} and with the outcome of our fits presented above. It is noteworthy that, at the lowest pion mass, the central value for $\langle x\rangle_{u-d}$ is {\em smaller}\, in the small volume $M_{\pi}L\sim 2.77$, compare Tab.~I of \cite{Bali:2014gha} (within error bars, the shift due to the different volumes is consistent with zero). In Fig.~\ref{fig:fvshifts}, we show the finite volume shifts $\delta\langle x\rangle_{u-d}:=\langle x\rangle_{u-d}^{L\rightarrow\infty}-\langle x\rangle_{u-d}(L)$ for three fixed values of $M_{\pi}L$, as a function of $M_{\pi}$, for the three fit scenarios $1\mathrm{fv}-3\mathrm{fv}$. As can be seen, these shifts scale as $\sim M_{\pi}^{2}$ for sufficiently small pion masses, as expected from the form of the loop corrections.
\begin{figure}
\centering
\subfigure[\,Fit scenario 1fv]{\includegraphics[width=0.42\textwidth]{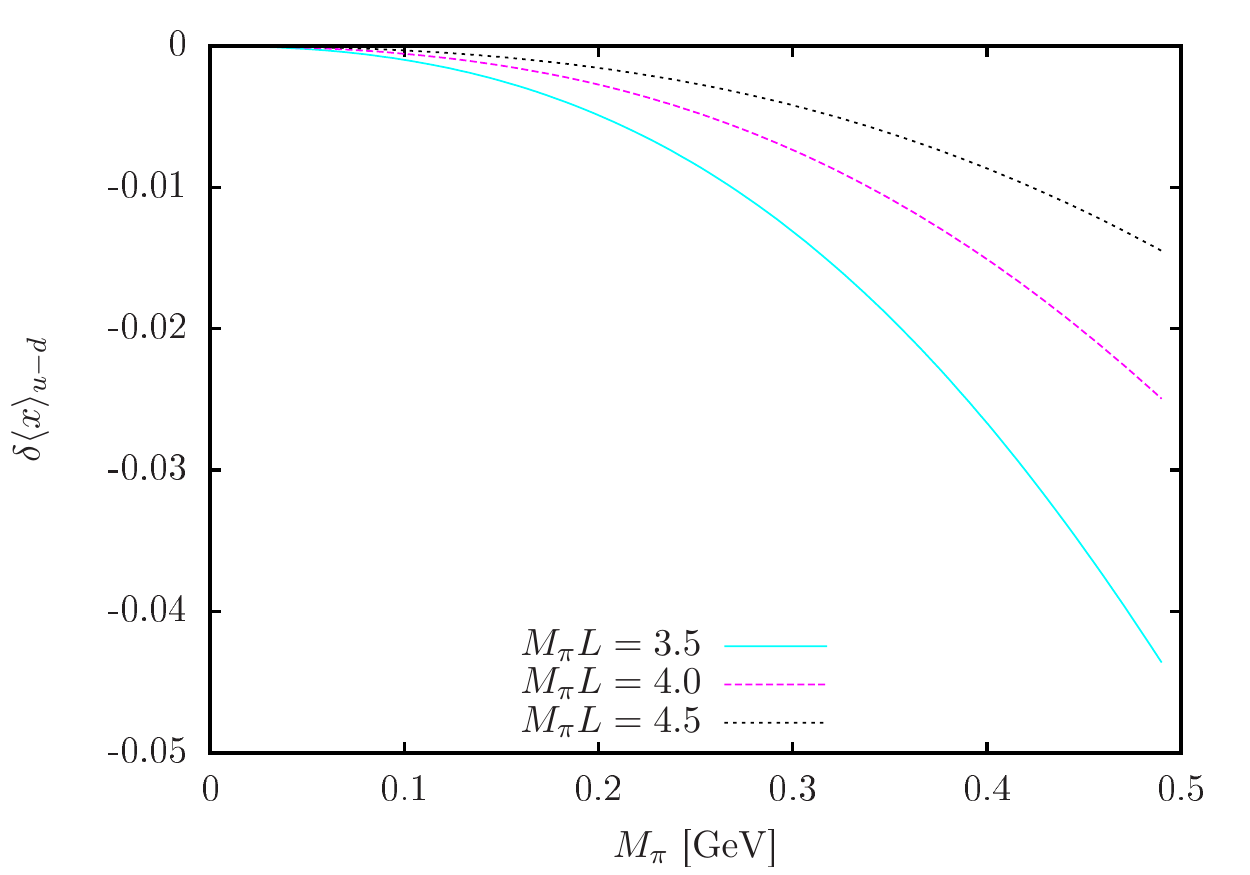}}
\subfigure[\,Fit scenario 2fv]{\includegraphics[width=0.42\textwidth]{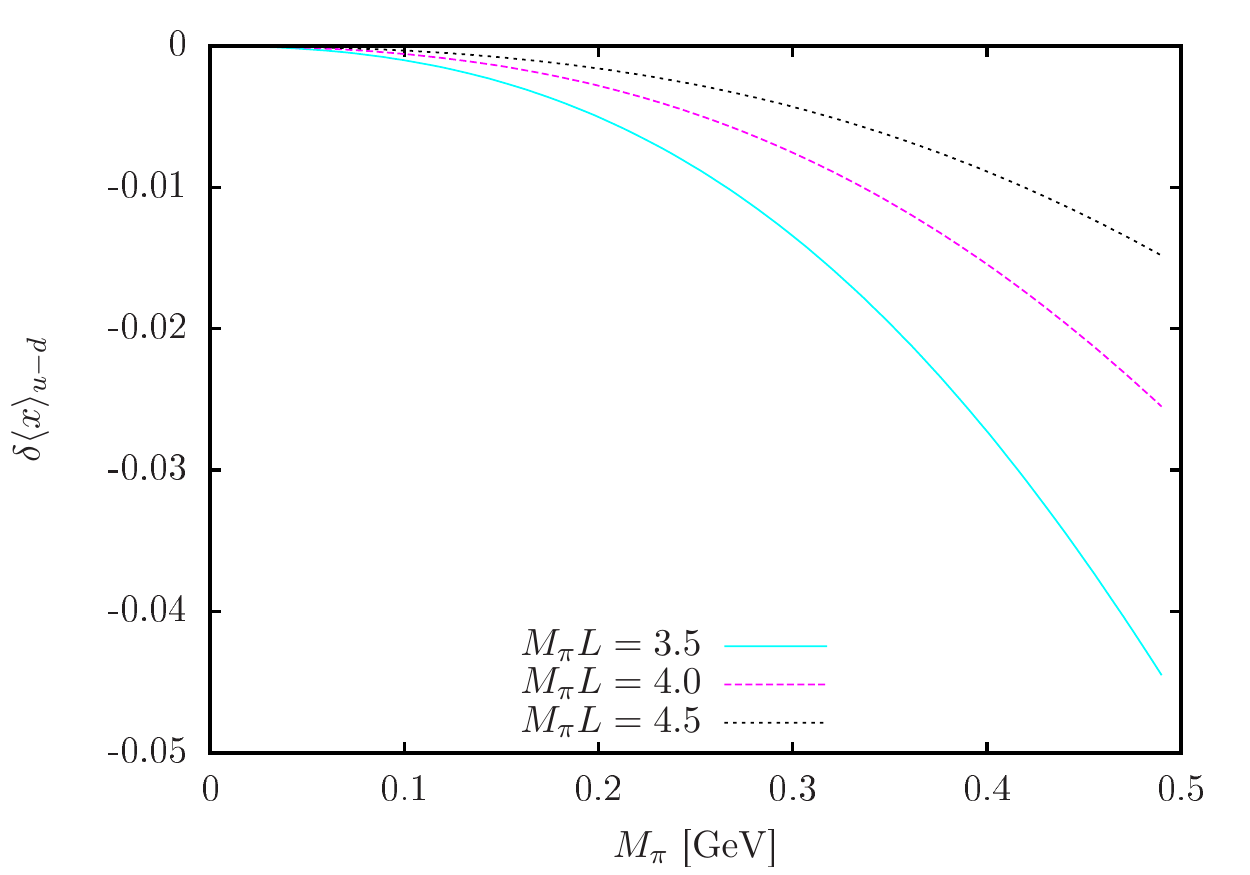}}
\subfigure[\,Fit scenario 3fv]{\includegraphics[width=0.42\textwidth]{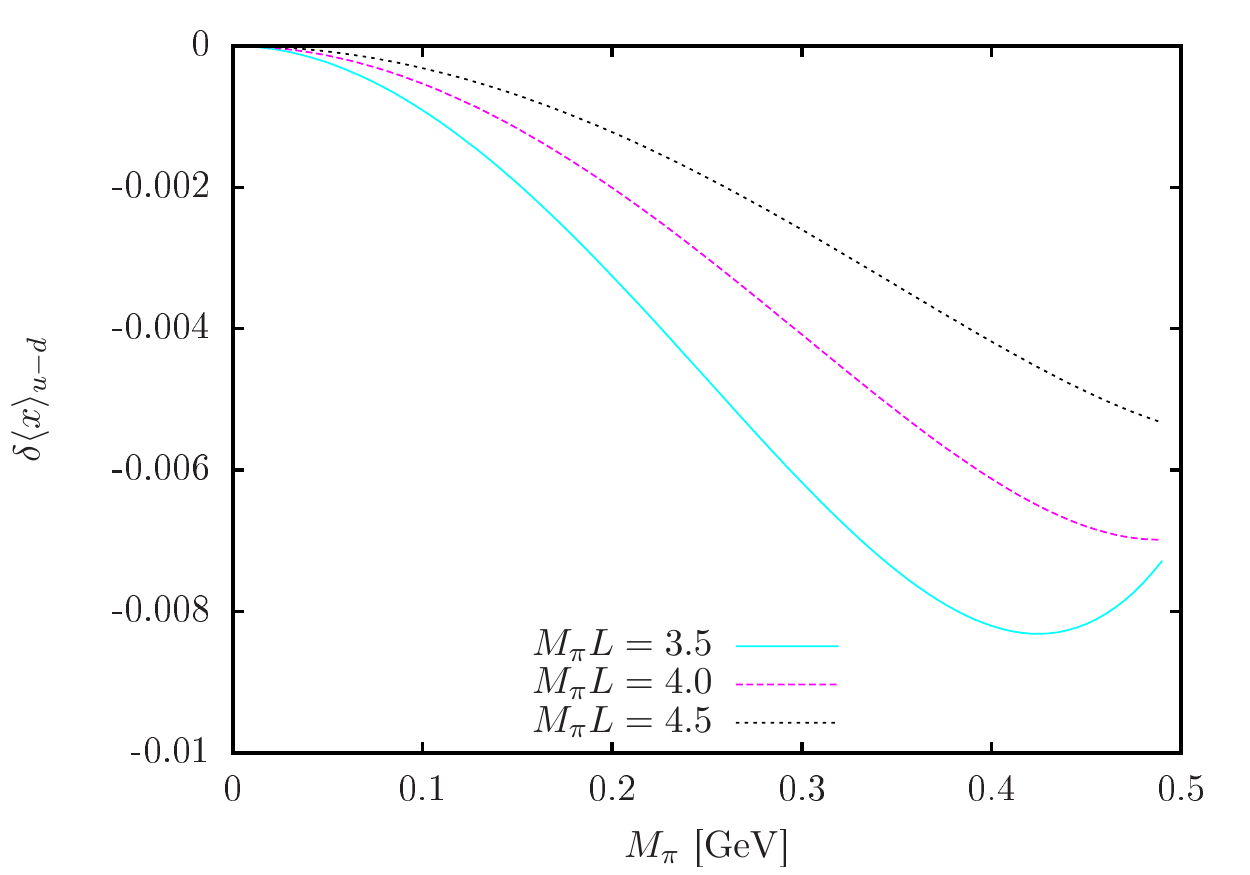}}
\caption{The finite volume shifts for fit scenarios $1-3$, for fixed $M_{\pi}L=3.5,\,4.0,\,4.5$.}
\label{fig:fvshifts}
\end{figure}%
\section{Discussion of results\label{sec:disc}}
\begin{figure}
\centering
\includegraphics[width=0.42\textwidth]{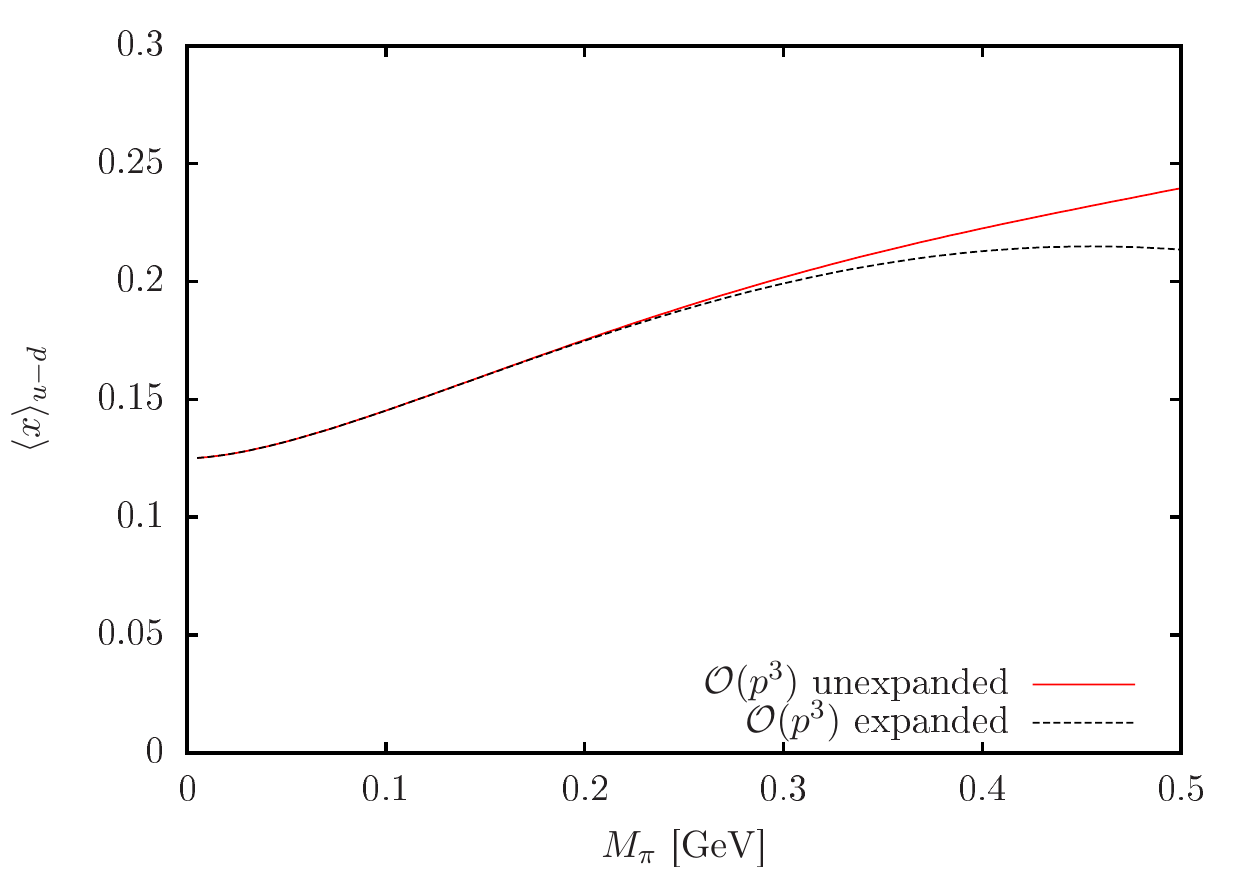}
\caption{The full one-loop expression for $\langle x\rangle_{u-d}$ in infinite volume (red), compared with Eq.~(\ref{eq:xumdexpansion}) (black, dashed), for the parameters of Eq.~(\ref{eq:parsfit2}) and $k_{i}=0$.}
\label{fig:fullvsexp}
\end{figure}

The three fit scenarios studied in Secs. \ref{sec:stab} and \ref{sec:finite} correspond to three different ways of using the BChPT formulae. From the point of view of lattice practitioners, our scenarios $1(\mathrm{fv})$ and $3(\mathrm{fv})$ correspond to the natural way the formulae are applied: Fitting to the lattice data input, it is checked whether the value of the observable at the experimental point can be predicted from the extrapolation. From our fit $3\mathrm{fv}$, where the full data set including the point at the nearly physical mass ($M_{\pi}\sim 150\,\mathrm{MeV}$) of \cite{Bali:2014gha} is inserted, we have to conclude that this is {\em not} possible. The chiral extrapolation together with the extrapolation to $L\rightarrow\infty$ is unable to reconcile the lattice data with the phenomenological value: The finite volume corrections are much too small to allow for a downward shift of the value at the lowest lattice pion mass, sufficient to come close to the experimental result. Excluding the low-mass point, we arrive at fit scenario $1\mathrm{fv}$, where it seems that the opposite conclusion is true - here, the extrapolation returns the phenomenological value to a satisfying degree of accuracy. This is the situation already 
encountered some years ago \cite{Edwards:2006qx,Dorati:2007bk,Dorati:2007pv,Detmold:2001jb,Detmold:2002nf,Detmold:2003rq,Wang:2010hp} and therefore nothing new (note, however, that previous applications of the BChPT framework to $\langle x\rangle_{u-d}$ were only accurate at $\mathcal{O}(p^2)$, since the $M_{\pi}^{3}$ term was either neglected or taken with a fixed coefficient $\sim\Delta a_{2,0}^{v},a_{2,0}^{v}$). There are pitfalls here, however: The fit result may depend considerably on data points for relatively high $M_{\pi}$, where the one-loop chiral representation is not reliable any more in a strict sense, as is shown e.g. by the broad band in fit $1$ of Fig.~\ref{fig:xumdbands1}, the convergence behavior illustrated in Fig.~\ref{fig:comp}, and the fit results for scenario $1'$ in Sec.~\ref{sec:stab}. Moreover, it is not clear whether the data points with $M_{\pi}L<4$ can be adequately described with a finite volume formula employing the $p$-regime counting. Nonetheless, it is interesting to see that the instability observed for fits $1,1'$ in Sec.~\ref{sec:stab} is eliminated in fit $1'\mathrm{fv}$, where the additional constraints on the finite volume behavior (as measured on the lattice) are built in. Needless to say that, for a more reliable extrapolation of lattice data, it is certainly necessary 
to include more data points from large volumes and $M_{\pi}\lesssim 350\,\mathrm{MeV}$. Concerning the second perspective mentioned in Sec.~\ref{sec:intro}, in view of an application of the BChPT extrapolation formulae for cases in which the experimental value is not (accurately) known, we would like to point out that BChPT can be useful in such a situation, even if the region where it can be reliably applied is more limited than previously thought. This is demonstrated by the comparison of our fit scenarios $1(\mathrm{fv})$ and $3(\mathrm{fv})$: in our opinion, it might be taken as an indication for a problem with some of the lattice data points that the chiral fit curves generically have the tendency to bend down at least below $M_{\pi} \lesssim 200\,\mathrm{MeV}$, a trend which can not be inferred from the lowest lattice data point. This indication is significant, given that the finite volume corrections are indeed small, and given that BChPT works at least close to the physical point, showing the suppression of higher orders in $M_{\pi}/m_{0}\sim\frac{1}{7}$ (modulo enhanced chiral logs) imposed by the chiral power counting (compare e.g.~Eq.~(\ref{eq:conv})). Under these conditions, it seems unavoidable that the extrapolation curve bends down appreciably\footnote{It is interesting to note that this strong down-bending of $\langle x\rangle_{u-d}(M_{\pi})$ is also seen in the Chiral Soliton Quark Model (CSQM) \cite{Diakonov:1985eg,Diakonov:1987ty}, see Fig.~3 of \cite{Wakamatsu:2006dy}. Up to a constant shift, the extrapolation curve in this model strongly resembles the expected behavior of the BChPT result (like our fits $1$,$2$). The same is true for the nucleon mass $m_{N}(M_{\pi})$ in the CSQM \cite{Goeke:2005fs}.} when approaching the region $M_{\pi} \lesssim 200\,\mathrm{MeV}$. If, on the other hand, the curve is forced to describe the lowest lattice point (fit $3(\mathrm{fv})$), the $\chi^2/\text{d.o.f.}$ value increases considerably, giving rise to the reasonable suspicion that the included data points are not compatible with the expected chiral extrapolation without taking into account further systematic corrections. 
For example, it has been pointed out in \cite{Bali:2014gha} that the effects due to the finite lattice spacing $a$ are not fully under control. Such effects are not grasped by our present extrapolation formula either, so this is certainly a reasonable direction of further investigation. Furthermore, it has been pointed out in \cite{Green:2011fg} that effects due to excited states tend to be larger for smaller pion masses. Considering the apparent problems due to the lowest-mass data point of \cite{Bali:2014gha}, we think that even though the problem of excited state contamination has been carefully studied in that work, one should thoroughly continue this route of investigation. Comparing with results for nearly-physical pion masses of other collaborations, we remark that ETMC \cite{Alexandrou:2013jsa} finds a value for $\langle x\rangle_{u-d}$ consistent with the one of \cite{Bali:2014gha}, 
while LHPC \cite{Green:2012ud} obtains a result {\em consistent}\, with the phenomenological value, though with less statistics than achieved in \cite{Bali:2014gha}.\\
From the point of view of effective field theory, the experimental result is just another data point in addition to the lattice data, with $M_{\pi}\sim 135\,\mathrm{MeV}$ and $L\rightarrow\infty$, and it is interesting in itself to study the functional form and properties of the chiral extrapolation and the values of the LECs using this experimental result as a constraint. The determination of the LECs can be useful in the study of other observables, or for a combined fit to several different nucleon structure properties. It also allows to assess the region of applicability of the BChPT formalism, which is also a much-debated topic in the literature on effective field theories \cite{Walker-Loud:2013yua,McGovern:1998tm,Young:2002ib,Bernard:2003rp,Beane:2004ks,Djukanovic:2006xc,McGovern:2006fm,Schindler:2006ha,Colangelo:2006mp,Bernard:2007zu,Hall:2010ai,Bali:2012qs,Bruns:2012eh,Beane:2014oea}. For our fit strategy $2$ corresponding to this philosophy (related to the first perspective on chiral extrapolations mentioned in Sec.~\ref{sec:intro}), we find a remarkable stability of the functional form of the extrapolation, compare e.g. the results of fits $2$ and $2'$ in Tab.~\ref{tab:fit2000}, and also the result of fit $2\mathrm{fv}$ and $2'\mathrm{fv}$ including finite volume corrections (Tab.~\ref{tab:fvfits}). In particular, the comparison shows that we could obtain a rough estimate of the finite volume effects already from the fits $2,2'$ of Sec.~\ref{sec:stab}, where only the largest volumes were included. The pion mass dependence is also under control for these fits, as can be seen from the comparison of the full one-loop form (for $L\rightarrow\infty$), employed in Sec.~\ref{sec:finite}, with the truncated expansion of Eq.~(\ref{eq:xumdexpansion}). This is illustrated in Fig.~\ref{fig:fullvsexp}, which demonstrates that higher-order terms of $\mathcal{O}(p^4)$ contained in the full unexpanded loop functions become important only for $M_{\pi}\gtrsim 400\,\mathrm{MeV}$ for these fit results.
But even though the results of the fits $2\mathrm{fv}$, $2'\mathrm{fv}$ seem very natural and reliable from the BChPT viewpoint, we cannot conclude that they yield the correct extrapolation function, unless we have a good argument why only the data points with $M_{\pi}\lesssim 200\,\mathrm{MeV}$ are afflicted with some significant systematic error.\\
In our analysis, we have applied the framework outlined in \cite{Dorati:2007bk,Wein:2014wma,Greil:2014awa}, where a field corres\-ponding to the $\Delta(1232)$ resonance is not included as an explicit degree of freedom. For studies where this resonance is included explicitly, see \cite{Detmold:2002nf,Detmold:2003rq,Wang:2010hp}. It is non-trivial to include the $\Delta(1232)$ in manifestly covariant BChPT due to problems with the power-counting scheme, and the presence of additional unphysical degrees of freedom in the covariant description of higher-spin fields ($s\geq 1$). For a discussion of these issues we refer to \cite{Bernard:2003xf,Hacker:2005fh,Pascalutsa:2005nd}, and references cited therein. Working at $\mathcal{O}(p^3)$, having three free fit parameters at hand, we expect that, for fixed delta-nucleon mass splitting and sufficiently small pion masses, the extrapolation would only be mildly different when including the $\Delta$ field, as the effects due to the resonance can mostly be absorbed in the coefficients of the local interaction terms. A conclusion pointing in this direction can also be inferred from the results of Ref.~\cite{Wang:2010hp}. But still, this point clearly deserves further study. 
\section{Conclusion\label{sec:conc}}
Baryon Chiral Perturbation Theory can be used to study the chiral extrapolation of $\langle x\rangle_{u-d}$, but it should be applied with great care. In our opinion, one should make sure that: (1) Higher order corrections with a reasonable strength do not alter the resulting extra\-po\-lation curve dramatically, (2) the convergence properties of the expanded extrapolation formula are at least roughly in accord with the expectations from chiral power counting, (3) the consideration of the finite volume effects do not lead to unreasonably large shifts of the fitted LECs, (4) if data with $M_{\pi}\gtrsim 500\,\mathrm{MeV}$ is included in the fit, the presence of these points has no big influence on the results, and (5) the resulting LECs are of natural size. Under these conditions, one can talk of a reliable extrapolation. If the according experimental result is accurately known, this extrapolation can be directly checked, which may serve as a further test whether the lattice data is afflicted with a systematic error not incorporated in the effective field theory. As was pointed out in the discussion in the previous section, this may be also possible if the experimental value is not available: If the chiral fit to the data set only returns results with a high $\chi^2/\text{d.o.f.}$ value, with an extrapolation function showing a bad convergence of the chiral expansion already at relatively small quark masses, and/or LECs of an unexpected size, these features can be seen as an indication (though of course not a proof) that some uncontrolled systematic error is still present in the lattice calculation. As should have become clear from our previous discussion, there is a clear indication that such an error source (like e.~g. discretization effects) is present in the case of lattice data for $\langle x\rangle_{u-d}$.
This indication can be further sharpened if some systematics is observed in the examination of this criterion (like in our comparison of fits $3(\mathrm{fv})$ and $1(\mathrm{fv})$, which provided evidence that the behavior of the data for very low $M_{\pi}$ is problematic from the ChPT perspective). In this way, chiral extrapolations can prove useful even if simulations are performed at nearly physical pion masses and large volumes. It is this kind of application of BChPT which we propose to consider in present-day and future lattice studies of $\langle x\rangle_{u-d}$ and other quantities parameterizing the structure of the nucleon.\\
As an extension of the extrapolation framework used here \cite{Dorati:2007bk,Wein:2014wma,Greil:2014awa}, one should consider the effects due to the $\Delta(1232)$ resonance along the lines of \cite{Bernard:2003xf,Hacker:2005fh,Pascalutsa:2005nd}. On a level of higher accuracy, also isospin-breaking corrections should be incorporated. Additionally, a combined fit of several nucleon structure functions should finally be undertaken. Only then, the full strength of (B)ChPT comes into play, yielding relations between different observables imposed by chiral symmetry (and other symmetries of the strong interaction).
Finally, we would like to mention that the analogues of the moment $\langle x\rangle_{u-d}$ for the full baryon ground-state octet (the form factors $A_{20}(t)$ at $t\rightarrow 0$, for all combinations of baryon states and flavor structures of the operator insertions) have also been calculated in (three-flavor) BChPT at leading one-loop order \cite{Bruns:2011sh,Shanahan:2013xw}.
\acknowledgments{We thank P.~Wein for discussions, and C.~Alexandrou for a useful communication. This work was supported by the Deutsche Forschungsgemeinschaft SFB/Transregio 55.}
\bibliographystyle{apsrev}

\end{document}